\documentclass[american,english]{iopart}
\usepackage[T1]{fontenc}
\usepackage[latin9]{inputenc}
\usepackage{refstyle}
\usepackage{amsbsy}
\usepackage{amstext}
\usepackage{graphicx}

\makeatletter

\AtBeginDocument{\providecommand\subsecref[1]{\ref{subsec:#1}}}
\AtBeginDocument{\providecommand\secref[1]{\ref{sec:#1}}}
\AtBeginDocument{\providecommand\eqref[1]{\ref{eq:#1}}}
\AtBeginDocument{\providecommand\figref[1]{\ref{fig:#1}}}
\AtBeginDocument{\providecommand\enuref[1]{\ref{enu:#1}}}
\RS@ifundefined{subsecref}
  {\newref{subsec}{name = \RSsectxt}}
  {}
\RS@ifundefined{thmref}
  {\def\RSthmtxt{theorem~}\newref{thm}{name = \RSthmtxt}}
  {}
\RS@ifundefined{lemref}
  {\def\RSlemtxt{lemma~}\newref{lem}{name = \RSlemtxt}}
  {}

\usepackage{iopams}
\usepackage{setstack}

\usepackage[numbers, sort&compress]{natbib}
\usepackage{iopams}
\usepackage{bbold}
\usepackage{prettyref}

\usepackage{feynmp-auto}
\usepackage{stackrel}

\newcommand{\newblock}{}

\makeatother

\usepackage{babel}
\begin{document}
\global\long\def\GamJ{\Gamma^{K}}%
\global\long\def\Gamc{\Gamma^{c}}%
\global\long\def\T{\mathrm{T}}%
\global\long\def\m{m}%
\global\long\def\v{v}%
\global\long\def\j{\boldsymbol{j}}%
\global\long\def\k{\boldsymbol{k}}%
\global\long\def\T{\mathrm{T}}%

\title{Diagrammatics of free energies with fixed variance for high-dimensional
data}
\author{Tobias Kühn}
\address{Institut de la Vision, Sorbonne Université, CNRS, INSERM, 17 rue Moreau,
75012 Paris, France}
\address{Institut für Physiologie, Universität Bern, Muesmattstr 27a, 3012
Bern, Switzerland}
\ead{tobias.kuehn@unibe.ch}
\begin{abstract}
Systems with many interacting stochastic constituents are fully characterized
by their free energy. Computing this quantity is therefore the objective
of various approaches, notably perturbative expansions, which are
applied in problems ranging from high-dimensional statistics to complex
systems. However, a lot of these techniques are complicated to apply
in practice because they lack a sufficient organization of the terms
of the perturbative series. In this manuscript, we tackle this problem
by using Feynman diagrams, extending a framework introduced earlier
to the case of free energies at fixed variances. This diagrammatics
do not require the theory to expand around to be Gaussian, which allows
its application to the free energy of a spin system studied to derive
message-passing algorithms by Maillard et al. 2019. We complete their
perturbative derivation of the free energy in the thermodynamic limit.
Furthermore, we derive resummations to estimate the entropies of poorly
sampled systems requiring only limited statistics and we revisit earlier
approaches to compute the free energy of the Ising model, revealing
new insights due to the extension of our framework to the free energy
at fixed variances. We expect our approach to be particularly useful
for problems of high-dimensional statistics, like matrix factorization,
and the study of complex networks.
\end{abstract}
\maketitle

\section{Introduction}

A lot of interesting and challenging problems in physics are composed
of multiple complicated, but well characterized components, many of
them interacting in a pairwise fashion. This concerns, for example,
spin systems (Ising, Potts, XY, Heisenberg, spherical spin,...), in
which the single components are connected by pairwise interactions.
While being quite restricted, this setting still gives rise to complex
behavior, studied notably in the field of spin glasses \cite{Mezard87,Charbonneau23_SpinGlassFarBeyond}.
Similarly, simple liquids consist of particles subjected to two-body
interactions and would otherwise form an ideal gas \cite{hansen2013theory}.
For both type of systems, expansions in the interactions have been
used to derive free energies and thereby characterize the behavior
of the corresponding systems. 

Generally, this setting in which the parameters are known and observables
are computed, in other words, the system's statistics is determined,
is known as forward (or direct) problem. In contrast to that, in praxis
one often knows the statistics of a complex systems, in many cases the
cumulants of its constituting single elements and the covariances
between them and one aims at computing the parameters of the underlying
system. This is known as inverse problem. There are plenty of examples
for data of this kind, for example electrophysiological recordings
from neuronal populations, measurements of the densities of animals
inside a flock or a herd or gene sequencing data. In none of these
cases, the underlying microscopic theory is known - or if it is, it
is not practically feasible to use it for predictions. Therefore,
effective models, borrowed from the microscopic description of spins,
are in use there. A popular and successful strategy in these examples
is to employ the reasoning of maximum entropy: given some statistical
measures, like, for example, the mean of a variable, one chooses among
all probability distributions reproducing them that one with the highest
entropy \cite{Jaynes03}. For binary variables with given means and
covariances, the resulting probability distribution is that of an
Ising spin glass, which is therefore a widespread tool in the modeling
of complex systems.

Both in forward and inverse problems, computing the free energies
of the respective systems means basically ``solving'' the system:
in the forward case, the free energy allows us to obtain all possible
moments by taking derivatives; in the inverse case, the same is true
for the parameters of the system. In the latter case, furthermore,
the free energy is equal to the maximum entropy, which is often interesting
on its own. 

In this study, we tackle the problem of computing the free energy
at fixed means and variances for both cases, that is, given interactions
and given covariances. For the forward problem, we will expand in
small couplings, for the inverse problem in small covariances. 

While these kind of expansions are not a novel concept, but can be
traced back at least to the seminal work of Plefka \cite{plefka1982convergence}
from 1982, on which follow-up work has built ever since \cite{georges1991expand,Parisi95_5267,Nakanishi97_8085,Sessak09,Cocco12_252,Jacquin16},
there are still problems that cannot be solved with the techniques
developed in these studies. Indeed, in a quite recent publication,
Maillard et al. \cite{Maillard19_113301} address the problem of deriving
the free energy of spherical spins coupled by a symmetric rotationally
invariant matrix. The use of the formalism of \cite{georges1991expand}
allows them to conjecture its form in the thermodynamic limit, but
not to give the complete proof. 

In this manuscript, we pick up these lines of thought, however, developing
them in terms of Feynman diagrams. As opposed to most other diagrammatic
frameworks, we will not assume the theory to expand about is Gaussian.
By extending the applicability of Feynman diagrammatics, we leverage
the strength of this concept - that it simplifies and structures computations
- to expansions around non-Gaussian theories. This is an important
addition because a number of models - like Ising or Heisenberg for
micromagnetism or simple liquids in soft-matter theory - is mostly
complicated because their core ingredients are non-Gaussian - even
though they are well tractable besides this. Overcoming the restriction
of diagrammatics to expansions around Gaussian theories allows us,
in particular, to complete the proof of Maillard et al, in \subsecref{Application_pSpins}.
Furthermore, we show, in \subsecref{Application_MaxEnt}, how to use
our framework in a inverse-problem setting to derive approximations
for the entropy of complex systems \cite{Kuehn25a_arxiv} and we discuss
how extending the Legendre transform to encompass the variance helps
to understand earlier results on perturbation theory for the Ising
model \cite{Kuehn18_375004,Cocco12_252,Vasiliev74,Vasiliev75} (\subsecref{Application_Ising}).

We derive the technical results on Feynman diagrammatics in \secref{Methods}
- for the expansion in small couplings in \subsecref{Methods_GamJ}
and for small covariances in \subsecref{Gaussian_LegTrafo} -, using
and extending techniques from our earlier works \cite{Kuehn18_375004,Kuehn23_115001}.
In the discussion, we give an outlook on possible applications of
our approach, including the derivation of message-passing algorithms
for matrix factorization \cite{Maillard22_083301}, the study of higher-order
correlations in complex systems and further insights into the diagrammatics
of the Ising model.

\section{Methods\label{sec:Methods}}

Consider a (Helmholtz) free energy of the form
\begin{equation}
\fl W\left[\boldsymbol{j},\boldsymbol{k}\right]=\ln\left(\sum_{\boldsymbol{\psi}}e^{-H\left[\boldsymbol{\psi}\right]+\sum_{i}\left(j_{i}\psi_{i}+k_{i}\psi_{i}^{2}\right)}\right),\label{eq:Def_cgf}
\end{equation}
where $\boldsymbol{\psi}$ is a discrete index. It could also be continuous,
in which case the sum over it would be an integral, a case however
that we will not consider in this manuscript. 

\subsection{The Legendre transform with respect to one-point source fields: $\protect\GamJ$\label{subsec:Methods_GamJ}}

Fixing the mean and the variance of the single spins, we obtain the
(Gibbs) free energy
\begin{equation}
\fl\GamJ\left[\boldsymbol{\m},\boldsymbol{\v}\right]:=\sup_{\boldsymbol{j},\boldsymbol{k}}\left[\sum_{i}\m_{i}j_{i}+\left(\v_{i}+m_{i}^{2}\right)k_{i}-W\left[\boldsymbol{j},\boldsymbol{k},K\right]\right]\label{eq:Def_Gamma_J}
\end{equation}
as a Legendre transform of \eqref{Def_cgf}. The supremum means that
we choose $\boldsymbol{j}$ and $\boldsymbol{k}$ such that
\begin{eqnarray*}
\fl m_{i} & = & \frac{\partial W}{\partial j_{i}}\\
\fl\left(\v_{i}+m_{i}^{2}\right) & = & \frac{\partial W}{\partial k_{i}},
\end{eqnarray*}
in other words: we choose the source fields such that the theory matches
the first two momenta $m_{i}$ and $v_{i}+m_{i}^{2}$. Differentiating
$\GamJ$ with respect to its arguments, we obtain the equations of
state
\begin{eqnarray}
\fl\frac{\partial\GamJ}{\partial m_{i}} & = & j_{i}+2m_{i}k_{i}\label{eq:EqOfState_GammaJ_delm}\\
\fl\frac{\partial\GamJ}{\partial v_{i}} & = & k_{i}\label{eq:EqOfState_GammaJ_delv}
\end{eqnarray}

\subsubsection{Derivation of the perturbative expansion for $\protect\GamJ$\label{subsec:Derivation_expansion_GamJ}}

Analogous to \cite{Kuehn18_375004,Kuehn23_115001}, we will rewrite
\eqref{Def_Gamma_J} in order to derive an expansion of $\GamJ$ in
the couplings $K_{ij}$, in which we structurally take into account
its definition as a Legendre transform. Necessarily, considerable
parts of what follows are paraphrases of \cite{Kuehn18_375004,Kuehn23_115001},
which we include to make this manuscript readable on its own. To prepare
the implementation of this expansion, we derive an expression for
$\GamJ$ without explicit reference to the Legendre transform as a
supremum. We use the definition of $\GamJ$, \eqref{Def_Gamma_J},
together with the equations of state, \eqref{EqOfState_GammaJ_delm}
and \eqref{EqOfState_GammaJ_delv}, through which we express $\boldsymbol{j}$
and $\boldsymbol{k}$, to arrive at

\begin{eqnarray*}
\fl &  & \sum_{\boldsymbol{\psi}}\exp\left(-H\left[\boldsymbol{\psi}\right]+j^{\T}\boldsymbol{\psi}+\frac{1}{2}\boldsymbol{\psi}^{\T}K\boldsymbol{\psi}\right)=\exp\left(W\left[j,K\right]\right)\\
\fl & = & \exp\left(\boldsymbol{\m}^{\T}\boldsymbol{j}+\left(\boldsymbol{\v}+\boldsymbol{m}^{2}\right)^{\T}\boldsymbol{k}-\GamJ\left[\boldsymbol{\m},\boldsymbol{\v}\right]\right)\\
\fl & = & \exp\left(\boldsymbol{\m}^{\T}\left(\frac{\partial\GamJ}{\partial\text{\ensuremath{\boldsymbol{\m}}}}-2\frac{\partial\GamJ}{\partial\boldsymbol{\v}}\odot\boldsymbol{\m}\right)+\left(\boldsymbol{\v}+\boldsymbol{\m}^{2}\right)^{\mathrm{T}}\frac{\partial\GamJ}{\partial\boldsymbol{\v}}-\GamJ\left[\boldsymbol{\m},\boldsymbol{\v}\right]\right),
\end{eqnarray*}
where by $\odot$ we denote element-wise multiplication. We isolate
$\GamJ$, which leads to

\begin{eqnarray}
\fl &  & \exp\left(-\GamJ\left[\boldsymbol{\m},\boldsymbol{\v}\right]\right)\nonumber \\
\fl & = & \sum_{\boldsymbol{\psi}}\,\exp\left(-H\left[\boldsymbol{\psi}\right]+\left(\boldsymbol{\psi}-\boldsymbol{m}\right)^{\T}\left(\frac{\partial\GamJ}{\partial\boldsymbol{m}}-2\frac{\partial\GamJ}{\partial\boldsymbol{v}}\odot\boldsymbol{m}\right)+\frac{\partial\GamJ}{\partial\boldsymbol{v}}\left[\left(\boldsymbol{\psi}^{2}-\boldsymbol{\m}^{2}\right)-\boldsymbol{v}\right]\right)\label{eq:Rewrite_GamJ_isolated}
\end{eqnarray}
Next, we will use this expression to set up the small-coupling expansion
for the quantity $\GamJ$. We separate the Hamiltonian and the free energies into a solvable
and a perturbing part, according to
\begin{eqnarray*}
\fl\GamJ & = & \GamJ_{0}+\GamJ_{V}
\end{eqnarray*}
and analogous for $H$ and $W$. Plugging this decomposition into
\eqref{Rewrite_GamJ_isolated}, isolating $-\GamJ_{V}$ on the left-hand
side, reuniting the $\GamJ_{0}$ with its derivatives to re-obtain
$W_{0}$, we get
\begin{eqnarray}
\fl &  & \exp\left(-\GamJ_{V}\left[\boldsymbol{\m},\boldsymbol{\v}\right]\right)\nonumber \\
\fl & = & e^{-W_{0}\left[\boldsymbol{j}_{0},\boldsymbol{k}_{0}\right]}\left.e^{-H_{V}\left[\frac{\partial}{\partial\boldsymbol{j}}\right]+\left(\frac{\partial\GamJ_{V}}{\partial\boldsymbol{m}}\right)^{\T}\left(\frac{\partial}{\partial\boldsymbol{j}}-\boldsymbol{m}\right)+\frac{\partial\GamJ_{V}}{\partial\boldsymbol{v}}\left[\left(\frac{\partial}{\partial\boldsymbol{j}}-\boldsymbol{\m}\right)^{2}-\boldsymbol{v}\right]}e^{W_{0}\left[\boldsymbol{j},\boldsymbol{k}_{0}\right]}\right|_{\boldsymbol{j}=\boldsymbol{j}_{0},\boldsymbol{k}=\boldsymbol{k}_{0}},\label{eq:Rewrite_GamJ_isolated-1}
\end{eqnarray}
where
\begin{eqnarray*}
\fl\boldsymbol{j}_{0} & = & \frac{\partial\GamJ_{0}}{\partial\boldsymbol{m}}-2\frac{\partial\GamJ_{0}}{\partial\boldsymbol{v}}\odot\boldsymbol{m}\\
\fl\boldsymbol{k}_{0} & = & \frac{\partial\GamJ_{0}}{\partial\boldsymbol{v}},
\end{eqnarray*}
in words, we choose the source fields such that the unperturbed theory
reproduces the measured statistics. As in \cite{Kuehn18_375004,Kuehn23_115001},
we define a differential operator for convenience, as 
\begin{eqnarray}
\fl A_{K}\left(\frac{\partial}{\partial\boldsymbol{j}}\right) & := & -H_{V}\left[\frac{\partial}{\partial\boldsymbol{j}}\right]\label{eq:A_operator_H}\\
\fl &  & +\left(\frac{\partial\GamJ_{V}}{\partial\boldsymbol{m}}\right)^{\T}\left(\frac{\partial}{\partial\boldsymbol{j}}-\boldsymbol{m}\right)\label{eq:A_operator_del_m}\\
\fl &  & +\left(\frac{\partial\GamJ_{V}}{\partial\boldsymbol{v}}\right)^{\T}\left[\left(\frac{\partial}{\partial\boldsymbol{j}}-\boldsymbol{\m}\right)^{2}-\boldsymbol{v}\right]\label{eq:A_operator_del_v}
\end{eqnarray}
and write 
\begin{equation}
\fl\exp\left(-\GamJ_{V}\left[\boldsymbol{\m},\boldsymbol{\v}\right]\right)=e^{-W_{0}\left[\boldsymbol{j}_{0},\boldsymbol{k}_{0}\right]}\left.\underset{L\rightarrow\infty}{\lim}\left(1+\frac{1}{L}A_{K}\left(\frac{\partial}{\partial\boldsymbol{j}}\right)\right)^{L}e^{W_{0}\left[\boldsymbol{j},\boldsymbol{k}\right]}\right|_{\boldsymbol{j}=\boldsymbol{j}_{0},\boldsymbol{k}=\boldsymbol{k}_{0}}\label{eq:Rewrite_GamJv_DefExp}
\end{equation}
so that we can compute $\GamJ_{V}\left[\boldsymbol{\m},\boldsymbol{\v}\right]$
as a series expansion by evaluating the perturbing terms order by
order in the interaction $K$. To organize the expansion in $K$,
we artificially discriminate between the $L$ in the denominator and
the $L$ in the power defining \cite{Kuehn18_375004,Kuehn23_115001}
\[
\fl e^{g_{l,L}\left[\boldsymbol{j},\boldsymbol{k}\right]}:=\left(1+\frac{1}{L}A_{K}\left(\frac{\partial}{\partial\boldsymbol{j}}\right)\right)^{l}e^{W_{0}\left[\boldsymbol{j},\boldsymbol{k}\right]}.
\]
We read off from \eqref{Rewrite_GamJv_DefExp} that
\[
\fl-\GamJ_{V}\left[\boldsymbol{\m},\boldsymbol{\v}\right]=-W_{0}\left[\boldsymbol{j}_{0},\boldsymbol{k}_{0}\right]+\lim_{L\rightarrow\infty}g_{L,L}\left[\boldsymbol{j}_{0},\boldsymbol{k}_{0}\right]=:-W_{0}\left[\boldsymbol{j}_{0},\boldsymbol{k}_{0}\right]+\lim_{L\rightarrow\infty}g\left[\boldsymbol{j}_{0},\boldsymbol{k}_{0}\right].
\]
By definition, we have
\[
\fl e^{g_{l+1,L}\left[\boldsymbol{j},\boldsymbol{k}\right]}=\left(1+\frac{1}{L}A_{K}\left(\frac{\partial}{\partial\boldsymbol{j}}\right)\right)e^{g_{l,L}\left[\boldsymbol{j},\boldsymbol{k}\right]},
\]
so that we obtain the recursion relation
\begin{eqnarray*}
\fl &  & g_{l+1,L}\left[\boldsymbol{j},\boldsymbol{k}\right]-g_{l,L}\left[\boldsymbol{j},\boldsymbol{k}\right]\\
\fl & = & -\frac{1}{L}e^{-g_{l,L}\left[\boldsymbol{j},\boldsymbol{k}\right]}H_{V}\left[\frac{\partial}{\partial\boldsymbol{j}}\right]e^{g_{l,L}\left[\boldsymbol{j},\boldsymbol{k}\right]}\\
\fl &  & +\frac{1}{L}e^{-g_{l,L}\left[\boldsymbol{j},\boldsymbol{k}\right]}\left(\frac{\partial\GamJ_{V}}{\partial\boldsymbol{m}}\right)^{\T}\left(\frac{\partial}{\partial\boldsymbol{j}}-\boldsymbol{m}\right)e^{g_{l,L}\left[\boldsymbol{j},\boldsymbol{k}\right]}\\
\fl &  & +\frac{1}{L}e^{-g_{l,L}\left[\boldsymbol{j},\boldsymbol{k}\right]}\left(\frac{\partial\GamJ_{V}}{\partial\boldsymbol{v}}\right)^{\T}\left[\left(\frac{\partial}{\partial\boldsymbol{j}}-\boldsymbol{\m}\right)^{2}-\boldsymbol{v}\right]e^{g_{l,L}\left[\boldsymbol{j},\boldsymbol{k}\right]}+\mathcal{O}\left(\frac{1}{L^{2}}\right).
\end{eqnarray*}
Note that in first order of the interaction, only $H_{V}\left[\frac{\partial}{\partial\boldsymbol{j}}\right]$
contributes, whereas the contributions of the other two parts cancel
by construction - as it has to be because of the properties of the
Legendre transform. We can therefore construct the small-$K$ expansion
of $\GamJ$ order by order, keeping track of all terms denoting them
by Feynman diagrams. Their constituting elements are

\begin{fmffile}{Collection_diagram_elements}	
	\begin{eqnarray}
		\fl
		\parbox[35mm]{25mm}{
			\begin{fmfgraph*}(25,25)
				\fmfpen{0.5thin}
				\fmftop{o1,o2,o3,o4,o5}
				\fmfbottom{u1,u2,u3,u4,u5}
				\fmf{phantom}{u1,m1,o1}
				\fmf{phantom}{u3,m3,o3}
				\fmf{phantom}{u5,m5,o5}
				\fmf{plain}{m3,o4}
				\fmf{plain}{m3,o5}
				\fmf{plain,tension=0}{m3,m5}
				\fmf{phantom, tension=10}{m1,m3}
				\fmf{plain,tension=2.5}{m3,u5}
				\fmfv{decor.shape=circle,decor.filled=empty, decor.size=6.5thin}{m3}
				\fmfv{label=$i$, label.angle=90, label.dist=4.5pt}{o4}
				\fmfv{label=$j$, label.angle=60, label.dist=2.pt}{o5}
				\fmfv{label=$j$, label.angle=20, label.dist=1.5pt}{m5}
				\fmfv{label=$...$, label.angle=0, label.dist=1.5pt}{u5}
			\end{fmfgraph*}
		}  \mkern-55mu = \left.\frac{\delta^{n} W}{\delta j_{i}\delta j_{j}\delta j_{k}...}\right|_{\boldsymbol{j} = \boldsymbol{j}_{0}, \boldsymbol{k} = \boldsymbol{k}_{0}} = \kappa^{n}_{i,j,k,...}
		= \mkern-15mu \parbox[35mm]{25mm}{
			\begin{fmfgraph*}(25,25)
				\fmfpen{0.5thin}
				\fmftop{o1,o2,o3,o4,o5}
				\fmfbottom{u1,u2,u3,u4,u5}
				\fmf{phantom}{u1,m1,o1}
				\fmf{phantom}{u3,m3,o3}
				\fmf{phantom}{u5,m5,o5}
				\fmf{plain}{m3,o4}
				\fmf{plain}{m3,o5}
				\fmf{plain,tension=0}{m3,m5}
				\fmf{phantom, tension=10}{m1,m3}
				\fmf{plain,tension=2.5}{m3,u5}
				\fmfv{decor.shape=circle,decor.filled=full, decor.size=6.5thin}{m3}
				\fmfv{label=$i$, label.angle=90, label.dist=4.5pt}{o4}
				\fmfv{label=$j$, label.angle=60, label.dist=2.pt}{o5}
				\fmfv{label=$k$, label.angle=20, label.dist=1.5pt}{m5}
				\fmfv{label=$...$, label.angle=0, label.dist=1.5pt}{u5}
			\end{fmfgraph*}
		}  \mkern-55mu ,
		\mkern 15mu
		\parbox{25mm}{
			\begin{fmfgraph*}(25,25)
				\fmfpen{0.5thin}
				\fmftop{o1,o2,o3}
				\fmfbottom{u1,u2,u3}
				\fmf{plain}{u1,o2}
				\fmf{plain}{u3,o2}
				\fmfv{label=$i$, label.angle=-90, label.dist=4.5pt}{u1}
				\fmfv{label=$j$, label.angle=-90, label.dist=4.5pt}{u3}
			\end{fmfgraph*}
		} \mkern-60mu = K_{ij}.\label{eq:Def_elements_diagrams} 
	\end{eqnarray}
\end{fmffile}

We have introduced two different symbols for the same class of objects,
the cumulants of the unperturbed theory, and will use either of them
depending on where they occur. When a cumulant appears as an element
of a derivative of $\GamJ_{v}$, we use full circles; when it occurs
as a derivative of $W_{0}$ evaluated at the source fields $\boldsymbol{j}_{0}$,
$\boldsymbol{k}_{0}$, we use empty circles. This allows us to identify
more quickly the origin of each part of a diagram.

\subsubsection{Diagrammatics of $\protect\GamJ$}

As for the first Legendre transform (with respect to $\boldsymbol{j}$
only) the first two orders of $\GamJ$ are given as

\begin{figure}
\includegraphics[width=0.9\textwidth]{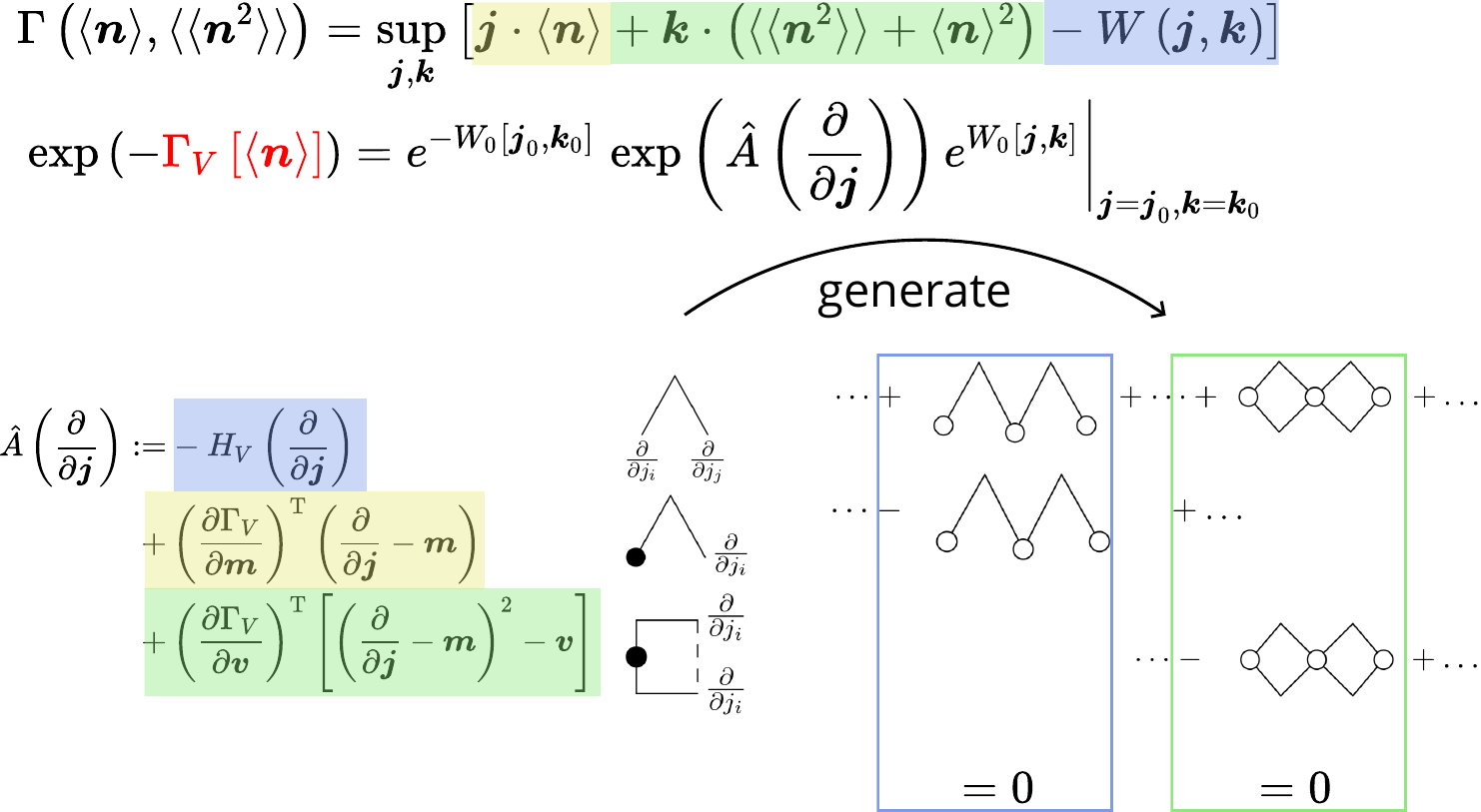}

\caption{Sketch of the cancellation mechanism of one-line reducible and cactus
diagrams.\label{fig:Sketch_cancellation_diagrams}}
\end{figure}

\begin{fmffile}{GammaJ_first_two}	
	\begin{equation}
		\fl \Gamma^{K}\left[\boldsymbol{m},\boldsymbol{v}\right] = \Gamma^{K}_{0}\left[\boldsymbol{m},\boldsymbol{v}\right] 
			- \mkern 20mu \parbox{25mm}{
				\begin{fmfgraph*}(25,25)
					\fmfpen{0.5thin}
					\fmftop{o1,o2,o3}
					\fmfbottom{u1,u2,u3}
					\fmf{plain}{u1,o2}
					\fmf{plain}{u3,o2}
					\fmfv{decor.shape=circle,decor.filled=empty, decor.size=6.5thin}{u1,u3}
				\end{fmfgraph*}
			}\mkern-60mu
			- \mkern-30mu 
			\parbox{25mm}{
				\begin{fmfgraph*}(75,25)
					\fmfpen{0.5thin}
					\fmftop{o1,o2,o3,o4,o5}
					\fmfbottom{u1,u2,u3,u4,u5}
					\fmf{phantom}{u1,v1,o3}
					\fmf{plain}{v1,o3}
					\fmf{phantom}{o1,v1,u3}
					\fmf{plain}{v1,u3}
					\fmf{phantom}{u3,v2,o5}
					\fmf{plain}{u3,v2}
					\fmf{phantom}{o3,v2,u5}
					\fmf{plain}{o3,v2}
					\fmfv{decor.shape=circle,decor.filled=empty, decor.size=6.5thin}{v1,v2}
				\end{fmfgraph*}
			} + \dots 
	\end{equation}
\end{fmffile}The contribution (\ref{eq:A_operator_H}) generates all connected
diagrams, according to the well-established linked-cluster theorem
\cite{ZinnJustin96}\cite[appendix A.3]{Kuehn18_375004}. Then, as
shown in \cite{Kuehn18_375004}, the contribution (\ref{eq:A_operator_del_m})
to the differential operator $A\left(\frac{\partial}{\partial\boldsymbol{j}}\right)$
removes all one-line reducible diagrams - that is, all diagrams from
the expansion of $W$  that can be split into two parts by cutting
a leg of an interaction (represented by an edge, cf. \figref{Sketch_cancellation_diagrams}).
In second order in $K$ for example, this means that the diagram

\begin{fmffile}{Three_cherries_diagram}	
	\begin{equation*}
		\fl \mkern 10mu\parbox{25mm}{
			\begin{fmfgraph*}(50,25)
				\fmfpen{0.5thin}
				\fmftop{o1,o2,o3,o4,o5}
				\fmfbottom{u1,u2,u3,u4,u5}
				\fmf{plain}{u1,o2}
				\fmf{plain}{o2,u3}
				\fmf{plain}{u3,o4}
				\fmf{plain}{o4,u5}
				\fmfv{decor.shape=circle,decor.filled=empty, decor.size=6.5thin}{u1,u3,u5}
			\end{fmfgraph*}
		} 
	\end{equation*}
\end{fmffile}is canceled in the expansion of $\GamJ$. In the following, we convince
ourselves that the effect of \eqref{A_operator_del_v} is analogous
and concerns two-particle reducible diagrams. For illustrative purposes,
we will start with the fourth order and then generalize. 

\subsubsection{Cancellations in fourth order\label{subsec:Cancellation_fourth_order_GamJ}}

The lowest-order contribution to \eqref{A_operator_del_v} is given
by

\begin{fmffile}{Two_point_int_from_v_deriv}
	\begin{eqnarray}
		\fl \frac{\partial \Gamma^{K}_{V}}{\partial v_i}
		\approx 
		-\mkern 20mu\parbox{20mm}{
			\begin{fmfgraph*}(20,30) 				
				\fmfpen{0.5thin} 				
				\fmftop{o1,o2} 				
				\fmfbottom{u1,u2}
				\fmf{plain}{o1,v1,u1}
				\fmf{plain}{o1,o2}
				\fmf{plain}{u1,u2}
				\fmf{dashes}{o2,u2}
				\fmfv{decor.shape=circle,decor.filled=full, decor.size=6.5thin}{v1}				 			
			\end{fmfgraph*} 		
		}
		\mkern-60mu
		= -\frac{\partial}{\partial v_{i}}\frac{1}{4}\sum_{k\neq j}v_{k}K_{kj}^{2}v_{j}=-\frac{1}{2}\sum_{j\left(\neq i\right)}K_{ij}^{2}v_{j},\label{eq:Interact_scd_order}
	\end{eqnarray}
\end{fmffile}where by the broken lines in diagrams we indicate identical indices
from now on. We recall from \eqref{Def_elements_diagrams} that full
nodes indicate cumulants, that is, the cumulant-generating functional
evaluated at $\boldsymbol{j}_{0}$ and $\boldsymbol{k}_{0}$. The
additional interaction in \eqref{Interact_scd_order} generates two
types of additional diagrams. First, there are the ones of the ``spectacles''
topology, which are depicted as the last two in the following equation.
For completeness, we add the diagram from the expansion of $W$ as
first diagram, which yields

\begin{fmffile}{Glasses_diagrams}
	\begin{eqnarray}
		\fl -
		\mkern 10mu
		\parbox{25mm}{
				\begin{fmfgraph*}(45,30)
					\fmfpen{0.5thin}
					\fmfleft{l1}
					\fmftop{o1,o2,o3,o4,o5}
					\fmfbottom{u1,u2,u3,u4,u5}
					\fmfright{r1}
					
					\fmf{phantom}{u3,v1,o3}					
	
					\fmf{plain}{l1,o2}
					\fmf{plain}{o2,v1}
					\fmf{plain}{v1,o4}
					\fmf{plain}{o4,r1}
					
					\fmf{plain}{l1,u2}					
					\fmf{plain}{u2,v1}
					\fmf{plain}{v1,u4}
					\fmf{plain}{u4,r1}

					\fmfv{label=$K$, label.angle=90, label.dist=2.5pt}{o2,o4}
					\fmfv{label=$K$, label.angle=-90, label.dist=2.5pt}{u2,u4}
														
					\fmfv{decor.shape=circle,decor.filled=empty, decor.size=6.5thin}{l1,v1,r1}
				\end{fmfgraph*}
		}
	\mkern-20mu + \mkern 50mu  
			\parbox{25mm}{
				\begin{fmfgraph*}(45,30)
					\fmfpen{0.5thin}
					\fmfleft{l1}
					\fmftop{o1,o2,o3,o4,o5}
					\fmfbottom{u1,u2,u3,u4,u5}
					\fmfright{r1}
					
					\fmf{phantom}{u3,v1,o3}					
	
					\fmf{plain}{l1,o2}
					\fmf{plain}{o2,v1}
					\fmf{plain}{v1,o4}
					\fmf{plain}{o4,r1}
					
					\fmf{plain}{l1,u2}					
					\fmf{plain}{u2,v1}
					\fmf{plain}{v1,u4}
					\fmf{plain}{u4,r1}
					
					\fmfv{decor.shape=circle,decor.filled=full, decor.size=6.5thin}{l1}
					\fmfv{label=$\frac{\partial \GamJ_{V}}{\partial \boldsymbol{v}}$, label.angle=180, label.dist=4.5pt}{l1}
					\fmfv{label=$K$, label.angle=90, label.dist=2.5pt}{o4}
					\fmfv{label=$K$, label.angle=-90, label.dist=2.5pt}{u4}
					\fmfv{decor.shape=circle,decor.filled=empty, decor.size=6.5thin}{v1,r1}
				\end{fmfgraph*}
		}
	\mkern-20mu - \mkern 50mu
		\parbox{25mm}{
				\begin{fmfgraph*}(45,30)
					\fmfpen{0.5thin}
					\fmfleft{l1}
					\fmftop{o1,o2,o3,o4,o5}
					\fmfbottom{u1,u2,u3,u4,u5}
					\fmfright{r1}
					
					\fmf{phantom}{u3,v1,o3}					
	
					\fmf{plain}{l1,o2}
					\fmf{plain}{o2,v1}
					\fmf{plain}{v1,o4}
					\fmf{plain}{o4,r1}
					
					\fmf{plain}{l1,u2}					
					\fmf{plain}{u2,v1}
					\fmf{plain}{v1,u4}
					\fmf{plain}{u4,r1}
														
					\fmfv{decor.shape=circle,decor.filled=full, decor.size=6.5thin}{l1,r1}
					\fmfv{label=$\frac{\partial \GamJ_{V}}{\partial \boldsymbol{v}}$, label.angle=180, label.dist=4.5pt}{l1}
					\fmfv{label=$\frac{\partial \GamJ_{V}}{\partial \boldsymbol{v}}$, label.angle=0, label.dist=4.5pt}{r1}
					\fmfv{decor.shape=circle,decor.filled=empty, decor.size=6.5thin}{v1}
				\end{fmfgraph*}
		} = 0\label{eq:Glasses_diagrams_fourth_order_GamJ}\\
		\nonumber	
	\end{eqnarray}
\end{fmffile}and we denote cumulants inside $\frac{\partial\GamJ_{V}}{\partial\boldsymbol{v}}$
vertices by full nodes to highlight their origin - as indicated in
\eqref{Def_elements_diagrams}, they translate into the same expressions
as the empty nodes, however. More generally, diagrams composed of
multiple rings attached at one point are known as cactus diagrams
\cite{Parisi95_5267,Maillard19_113301,Pappalardi22_170603}. The spectacles
diagrams come about by letting both derivatives associated to the
$\frac{\partial\GamJ_{_{V}}}{\partial\boldsymbol{v}}$ vertex act
on the same cumulant. The other possibility is to generate two cumulants
of order two. This yields, first,

\begin{fmffile}{Ring_diagrams_fourth_order_one_pair_identical}
	\begin{eqnarray}
		\fl \mkern 20mu
		- \mkern 10mu 
		\parbox{25mm}{
			\begin{fmfgraph*}(40,30) 				
				\fmfpen{0.5thin} 				
				\fmftop{o1,o2,o3} 				
				\fmfbottom{u1,u2,u3}
				\fmf{plain}{o1,v1,u1}
				\fmfleft{l1}
				\fmfright{r1}
				\fmf{dots, tension = 0}{l1,r1}
				\fmf{plain}{o3,v2,u3}
				\fmf{plain}{o1,o2,o3}
				\fmf{plain}{u1,u2,u3}
				\fmf{dashes, tension = 100}{o2,u2}
				\fmfv{decor.shape=circle,decor.filled=empty, decor.size=6.5thin}{l1,r1,o2,u2}
				\fmfv{label=$K$, label.angle=90, label.dist=2.5pt}{o1,o3}
				\fmfv{label=$K$, label.angle=-90, label.dist=2.5pt}{u1,u3}				 			
			\end{fmfgraph*} 		
		}
		\mkern-30mu + \mkern 45mu
		\parbox{25mm}{
			\begin{fmfgraph*}(40,30) 				
				\fmfpen{0.5thin} 				
				\fmftop{o1,o2,o3} 				
				\fmfbottom{u1,u2,u3}
				\fmf{plain}{o1,v1,u1}
				\fmfleft{l1}
				\fmfright{r1}
				\fmf{dots, tension = 0}{l1,r1}
				\fmf{plain}{o3,v2,u3}
				\fmf{plain}{o1,o2,o3}
				\fmf{plain}{u1,u2,u3}
				\fmf{dashes, tension = 100}{o2,u2}
				\fmfv{decor.shape=circle,decor.filled=empty, decor.size=6.5thin}{r1,o2,u2}
				\fmfv{decor.shape=circle,decor.filled=full, decor.size=6.5thin}{l1}
				\fmfv{label=$K$, label.angle=90, label.dist=2.5pt}{o3}
				\fmfv{label=$K$, label.angle=-90, label.dist=2.5pt}{u3}				 			
				\fmfv{label=$\frac{\partial \GamJ_{V}}{\partial \boldsymbol{v}}$, label.angle=180, label.dist=4.5pt}{l1}
			\end{fmfgraph*} 		
		}
		\mkern-30mu - \mkern 45mu
		\parbox{25mm}{
			\begin{fmfgraph*}(40,30) 				
				\fmfpen{0.5thin} 				
				\fmftop{o1,o2,o3} 				
				\fmfbottom{u1,u2,u3}
				\fmfleft{l1}
				\fmfright{r1}
				\fmf{dots, tension = 0}{l1,r1}
				\fmf{plain}{o1,l1,u1}
				\fmf{plain}{o3,r1,u3}
				\fmf{plain}{o1,o2,o3}
				\fmf{plain}{u1,u2,u3}
				\fmf{dashes, tension = 100}{o2,u2}
				\fmfv{decor.shape=circle,decor.filled=empty, decor.size=6.5thin}{o2,u2}
				\fmfv{decor.shape=circle,decor.filled=full, decor.size=6.5thin}{l1,r1}
				\fmfv{label=$\frac{\partial \GamJ_{V}}{\partial \boldsymbol{v}}$, label.angle=180, label.dist=4.5pt}{l1}
				\fmfv{label=$\frac{\partial \GamJ_{V}}{\partial \boldsymbol{v}}$, label.angle=0, label.dist=4.5pt}{r1}
			\end{fmfgraph*} 		
		}
		\mkern-20mu= 0,\label{eq:Ring_diagrams_fourth_order_GamJ_one_pair_equal}\\
		\nonumber
	\end{eqnarray}
\end{fmffile}where dotted lines denote pairs of nodes with indices that are explicitly
unequal (so this is the opposite of dashed lines). Complementary to
that, we have

\begin{fmffile}{Ring_diagrams_fourth_order_two_pairs_identical}
	\begin{eqnarray}
		\fl \mkern 20mu
		- \mkern 10mu 
		\parbox{25mm}{
			\begin{fmfgraph*}(40,30) 				
				\fmfpen{0.5thin} 				
				\fmftop{o1,o2,o3} 				
				\fmfbottom{u1,u2,u3}
				\fmf{plain}{o1,v1,u1}
				\fmfleft{l1}
				\fmfright{r1}
				\fmf{dashes, tension = 0}{l1,r1}
				\fmf{plain}{o3,v2,u3}
				\fmf{plain}{o1,o2,o3}
				\fmf{plain}{u1,u2,u3}
				\fmf{dashes, tension = 100}{o2,u2}
				\fmfv{decor.shape=circle,decor.filled=empty, decor.size=6.5thin}{l1,r1,o2,u2}
				\fmfv{label=$K$, label.angle=90, label.dist=2.5pt}{o1,o3}
				\fmfv{label=$K$, label.angle=-90, label.dist=2.5pt}{u1,u3}				 			
			\end{fmfgraph*} 		
		}
		\mkern-30mu + \mkern 45mu
		\parbox{25mm}{
			\begin{fmfgraph*}(40,30) 				
				\fmfpen{0.5thin} 				
				\fmftop{o1,o2,o3} 				
				\fmfbottom{u1,u2,u3}
				\fmf{plain}{o1,v1,u1}
				\fmfleft{l1}
				\fmfright{r1}
				\fmf{dashes, tension = 0}{l1,r1}
				\fmf{plain}{o3,v2,u3}
				\fmf{plain}{o1,o2,o3}
				\fmf{plain}{u1,u2,u3}
				\fmf{dashes, tension = 100}{o2,u2}
				\fmfv{decor.shape=circle,decor.filled=empty, decor.size=6.5thin}{r1,o2,u2}
				\fmfv{decor.shape=circle,decor.filled=full, decor.size=6.5thin}{l1}
				\fmfv{label=$K$, label.angle=90, label.dist=2.5pt}{o3}
				\fmfv{label=$K$, label.angle=-90, label.dist=2.5pt}{u3}				 			
				\fmfv{label=$\frac{\partial \GamJ_{V}}{\partial \boldsymbol{v}}$, label.angle=180, label.dist=4.5pt}{l1}
			\end{fmfgraph*} 		
		}
		\mkern-30mu - \mkern 45mu
		\parbox{25mm}{
			\begin{fmfgraph*}(40,30) 				
				\fmfpen{0.5thin} 				
				\fmftop{o1,o2,o3} 				
				\fmfbottom{u1,u2,u3}
				\fmfleft{l1}
				\fmfright{r1}
				\fmf{dashes, tension = 0}{l1,r1}
				\fmf{plain}{o1,l1,u1}
				\fmf{plain}{o3,r1,u3}
				\fmf{plain}{o1,o2,o3}
				\fmf{plain}{u1,u2,u3}
				\fmf{dashes, tension = 100}{o2,u2}
				\fmfv{decor.shape=circle,decor.filled=empty, decor.size=6.5thin}{o2,u2}
				\fmfv{decor.shape=circle,decor.filled=full, decor.size=6.5thin}{l1,r1}
				\fmfv{label=$\frac{\partial \GamJ_{V}}{\partial \boldsymbol{v}}$, label.angle=180, label.dist=4.5pt}{l1}
				\fmfv{label=$\frac{\partial \GamJ_{V}}{\partial \boldsymbol{v}}$, label.angle=0, label.dist=4.5pt}{r1}
			\end{fmfgraph*} 		
		}
		\mkern-20mu\neq 0.\label{eq:Ring_diagrams_fourth_order_GamJ_two_pairs_equal}\\
		\nonumber
	\end{eqnarray}
\end{fmffile}

By foreshadowing that all contributions in \eqref{Ring_diagrams_fourth_order_GamJ_one_pair_equal}
taken together yield $0$, we have implied that the symmetry factors
are the same, no matter if we compose the diagrams only with interactions
$K$, with elements of $\frac{\partial\GamJ_{V}}{\partial\boldsymbol{v}}$
or a mixture of both, and that only their signs change. For order
four, it is straight-forward to convince ourselves explicitly that
this is true. 

Indeed, the ring diagrams in \eqref{Ring_diagrams_fourth_order_GamJ_one_pair_equal}
translate to
\begin{eqnarray}
\fl &  & -\left(\frac{1}{4!}\frac{1}{2^{4}}\right)\cdot\left(\frac{4!}{4\cdot2}\cdot2^{4}\right)\cdot2\sum_{i\neq j\neq k\neq i}v_{j}K_{ij}^{2}v_{i}^{2}K_{ik}^{2}v_{k}\label{eq:RingDiagrams_GammaJ_FirstLine}\\
\fl &  & -\left(\frac{1}{2}\frac{1}{2^{2}}\right)\cdot\left(2^{2}\cdot2\right)\sum_{i\neq j}\frac{\partial\GamJ_{V}}{\partial v_{i}}v_{i}^{2}K_{ij}^{2}v_{j}-\frac{1}{2}\cdot2\sum_{i}\frac{\partial\GamJ_{V}}{\partial v_{i}}v_{i}^{2}\frac{\partial\GamJ_{V}}{\partial v_{i}}\label{eq:RingDiagrams_GammaJ_SecondLine}\\
\fl & = & -\frac{1}{4}\sum_{i\neq j\neq k\neq i}K_{ij}^{2}v_{i}^{2}K_{ik}^{2}+\frac{1}{2}\sum_{i\neq j}\sum_{k\neq i}v_{k}K_{ik}^{2}v_{i}^{2}K_{ij}^{2}v_{j}-\frac{1}{4}\sum_{i}\sum_{k\neq i}v_{k}K_{ik}^{2}v_{i}^{2}\sum_{j\neq i}v_{j}K_{ij}^{2}=0,\nonumber 
\end{eqnarray}
where we indicate the origin of the prefactors of the terms in \eqref{RingDiagrams_GammaJ_FirstLine}
and \eqref{RingDiagrams_GammaJ_SecondLine}, which is coming about
by the well-known Feynman rules described, e.g., in \cite[sec. 2.4.3.]{Kuehn18_375004}
and which we recall for this example : 
\begin{itemize}
\item The ring diagrams composed of $K$s, in \eqref{RingDiagrams_GammaJ_FirstLine},
contributes with a prefactor $\frac{1}{4!2^{4}}$, the $4!$ coming
from the Taylor expansion in small couplings, the $\frac{1}{2^{4}}$
from the fact that each of them contributes with a factor $\frac{1}{2}$.
We recall that to determine the symmetry factor of a diagram, we label
the legs of its vertices by indices and count the number of symmetry
operations (permutation of vertices and flipping the legs of a vertex)
that generate new combinations of labels, that is, new labeled diagrams.
There are $4!$ ways to permute the four interactions, but $4\cdot2$
of them lead to the same labeled diagram because of the symmetry of
the ring: one can start counting at every node and either go clockwise
or counter-clockwise. Then, all $2^{4}$ possible vertex flips generate
new labeled diagrams, so that the symmetry factor is as indicated
in the second bracket. Finally, there is another factor of $2$ only
occurring in the translation of \eqref{Ring_diagrams_fourth_order_GamJ_one_pair_equal},
but not of \eqref{Ring_diagrams_fourth_order_GamJ_two_pairs_equal},
which we explain below.
\item In the first term of \eqref{RingDiagrams_GammaJ_SecondLine}, we have
the prefactor $\frac{1}{2}$ because of the second order in $K$ and
$\frac{1}{2^{2}}$ because every $K$ comes with an $\frac{1}{2}$
and the symmetry factor is $2^{2}\cdot2$ because flipping either
of the $K$'s or the $\frac{\partial\GamJ_{V}}{\partial\boldsymbol{v}}$-vertex
generates a new labeled diagram.
\item In the second term of \eqref{RingDiagrams_GammaJ_SecondLine}, we
again have a prefactor $\frac{1}{2}$ because of the double occurrence
of a vertex, this time the one from $\frac{\partial\GamJ_{V}}{\partial\boldsymbol{v}}$
and the symmetry factor is $2$ because we can flip one of them to
generate a new labeled diagram (but flipping both brings back the
original one).
\end{itemize}
Finally, we come to the point not related to the Feynman rules proper,
which can only occur in case of discrete indices, namely the difference
between \eqref{Ring_diagrams_fourth_order_GamJ_one_pair_equal} and
\eqref{Ring_diagrams_fourth_order_GamJ_two_pairs_equal}. Indeed,
their respective first diagrams translate differently because in \eqref{Ring_diagrams_fourth_order_GamJ_one_pair_equal},
there are two possibilities to choose the pair of nodes bearing the
same index, whereas in \eqref{Ring_diagrams_fourth_order_GamJ_two_pairs_equal},
there is no such choice. In other words: in \eqref{Ring_diagrams_fourth_order_GamJ_one_pair_equal},
all diagrams have the same symmetry, whereas in \eqref{Ring_diagrams_fourth_order_GamJ_two_pairs_equal},
the symmetry is higher in the first contribution. Consequently, the
last factor $2$ in \eqref{RingDiagrams_GammaJ_FirstLine} occurs
only in the translation of \eqref{Ring_diagrams_fourth_order_GamJ_one_pair_equal}
and therefore, the sum of all diagrams indeed vanishes only there.
We will discuss this issue in more detail in the next section, \subsecref{Cancellation_higher_orders}.

Summarizing, the contribution of all ring diagrams of order four only
contains the contributions with four pairwise unequal indices and
the contribution with two pairs of nodes with identical indices:
\[
-\frac{1}{8}\sum_{\underset{i\neq k,\,j\neq l}{i\neq j\neq k\neq l\neq i}}v_{i}K_{ij}v_{j}K_{jk}v_{k}K_{kl}v_{l}K_{li}+\frac{1}{8}\sum_{i\neq j}v_{i}^{4}K_{ij}^{4}v_{j}^{4}.
\]

The reasoning for the ``spectacles'' diagram, as shown in \eqref{Glasses_diagrams_fourth_order_GamJ}
is analogous. The diagrams including $\frac{\partial\Gamma}{\partial v_{i}}$
come with the very same prefactors as the contributions before, only
divided by $2$ because they are no two nodes to choose from.\footnote{It might appear a bit odd at first sight that there is difference
between the two cases because topologically the two nodes of order
$2$ play the same role as that of order $4$. However, consider the
origin of the prefactors for, say, the respective last diagrams \eqref{Glasses_diagrams_fourth_order_GamJ}
and \eqref{Ring_diagrams_fourth_order_GamJ_one_pair_equal}: they
come about by letting differential operators act on $W_{0}$. In case
of the ring diagrams, this means the following: There are first two
cumulants of order $1$ each are generated by letting the derivatives
associated to one $\frac{\partial\GamJ_{V}}{\partial\boldsymbol{v}}$
vertex act on $W_{0}$. Then, letting the other $\frac{\partial\GamJ_{V}}{\partial\boldsymbol{v}}$
vertex act on these two cumulants there are always two ways of picking
one of them. In contrast, for the four-point cumulant, one does not
have this choice. In other words, when it comes to symmetry factors
the two cumulants of order $2$ are different than one cumulant of
order $4$.}. Correspondingly, also the spectacles diagram composed just of $K$s
comes with a prefactor of $\frac{1}{8}$ instead of $\frac{1}{4}$
- indeed we have the same factor of $\frac{1}{4!2^{4}}$ as for the
ring diagram, $3$ possibilities to pair vertices on either side and
$2^{4}$ vertex flips generating new diagrams. Therefore, we have
confirmed that also this contribution is canceled.

\subsubsection{Cancellations in higher orders\label{subsec:Cancellation_higher_orders}}

\begin{figure}
\includegraphics{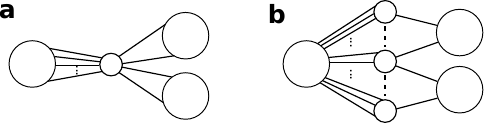}\caption{\label{fig:Cactus_and_PseudoCactus_diagrams}Sketch of (a) a cactus
diagram and (b) an improper pseudo-cactus diagram.}
\end{figure}

We now proceed to higher orders, leveraging the ideas employed for
the second-order terms. To do this, we introduce the notion of a counter
diagram: it has the same form as a diagram in the diagrammatic expansion
of $W$, but it contributes with the opposite sign. We will see that
there are counter diagrams for two types of diagrams, defined by consisting
of 
\begin{enumerate}
\item multiple sub-diagrams connected only by a single node and at least
one of them is connected to the node by exactly two lines. Generalizing
this notion from the literature, we call these diagrams cactus diagrams
(\figref{Cactus_and_PseudoCactus_diagrams}a)\label{enu:Definition_cactus}
\item multiple sub-diagrams connected only by nodes joined by dashed lines
and at least one of the sub-diagrams is connected to the two nodes
by exactly two lines (each of the two lines ending in a different
node). Henceforth, we call collections of nodes associated with identical
indices (indicated by dashed lines) pseudo-nodes and the corresponding
diagrams pseudo-cactus diagrams (\figref{Cactus_and_PseudoCactus_diagrams}b).\label{enu:Definition_pseudo_cactus}
\end{enumerate}
The notion of counter diagrams is similar to that of compensating
diagrams of Vasiliev and Radzhabov, introduced in the study of the
diagrammatics of the Ising model \cite{Vasiliev75}. We discuss this
point in a bit more detail in \secref{Discussion} and leave a more
thorough exploration of this link to future work.

Letting $\frac{\partial}{\partial v_{i}}$ act on that part of $\GamJ_{V}$
containing a second-order cumulant, this yields\begin{fmffile}{TwoPointInt_from_vDeriv_General}
	\begin{eqnarray}
		\fl \frac{\partial \GamJ_{V}}{\partial v_i} \left[\left(\frac{\partial}{\partial j_{i}}-m_{i}\right)^{2}-v_{i}\right] \
		\sim 
		-\mkern 20mu\parbox{20mm}{
			\begin{fmfgraph*}(20,30) 				
				\fmfpen{0.5thin} 				
				\fmftop{o1,o2} 				
				\fmfbottom{u1,u2}
				\fmf{plain}{o1,v1,u1}
				\fmf{plain}{o1,o2}
				\fmf{plain}{u1,u2}
				\fmf{dashes}{o2,u2}
				\fmfv{decor.shape=circle,decor.filled=shaded, decor.size=10.5thin}{v1}
				\fmfv{label=$\partial_{j_{i}}$, label.angle=0, label.dist=4.5pt}{o2,u2}				 			
			\end{fmfgraph*} 		
		}\mkern -36mu.
	\end{eqnarray}
\end{fmffile}Labeling the diagram with derivatives with respect to $j_{i}$, we
have taken into account that the contributions $-m_{i}$ and $-v_{i}$
on the left hand side just remove trivial contributions, in which
the sub-diagram is attached to isolated cumlants of order one or two,
the right-hand side therefore indicates how the left-hand side acts
as an operator. That is why we write $\sim$ instead of $=$. The
$\partial_{j_{i}}$ operators, in turn, can either act on the same
cumulant or on two different ones, corresponding to the points (\enuref{Definition_cactus})
and (\enuref{Definition_pseudo_cactus}) above, respectively. For
the fourth order, we have already shown that this leads to cancellations
because the new diagrams contribute with the same prefactors. We convince
ourselves that the same is true in the general case. As already argued
in \cite[sec. 2.4.1.]{Kuehn18_375004}, it is enough to compare the
prefactors of the subdiagrams. Indeed, when considering the diagrams
only constructed from $K_{ij}$, say, of order $n$, with $k$ interactions
in one sub-diagram and $n-k$ interactions in the other part, we can
determine its overall symmetry factor by first counting the ways one
can distribute the interactions on the two subdiagrams: there are
$\left(\begin{array}{c}
n\\
k
\end{array}\right)$ possibilities to do so. This is combined with the overall prefactor
of the diagram to yield $\left(\begin{array}{c}
n\\
k
\end{array}\right)\frac{1}{n!}=\frac{1}{k!\left(n-k\right)!},$ the same one we obtain from assembling the sub-diagrams and adding
them afterwards. The only aspect that is different than in \cite{Kuehn18_375004}
is that we stick together the subdiagrams at two legs, instead of
one. The resulting differences in the symmetry factors are the same,
no matter if we construct the diagrams with $\frac{\partial\GamJ_{V}}{\partial\boldsymbol{v}}$
vertices or with $K$s. 

We now address the question in which cases the existence of counter
diagrams leads to a cancellation. In fact, this is always the case
for the diagrams described in point 1 above, but not necessarily for
the diagrams in point (\enuref{Definition_pseudo_cactus}).

\begin{figure}
\includegraphics[width=1\textwidth]{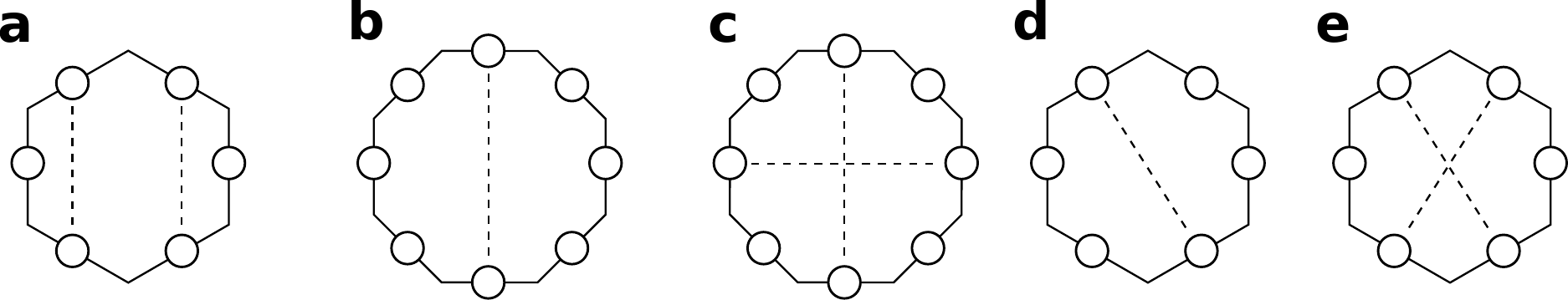}\caption{Ring diagrams with possible pairs of nodes with identical indices
indicated by dashed lines.\label{fig:Identifications_RingDiagrams_examples}}

\end{figure}

We will discuss this issue by considering ring diagrams - because
they are the simplest examples and also the most relevant for applications.
By treating all indices on the same footing (not taking into account
any possibilities of identifications), the prefactor of a ring diagrams
of order $n$ is $\frac{1}{n!}\frac{1}{2^{n}}\frac{n!}{2n}2^{n}=\frac{1}{2n}$.
Indeed, every vertex flip generates a new labeled diagram, but not
every permutation of vertices because one can start counting them
at any node and label them clockwise or counter-clockwise.

We first study the ring diagram of sixth order in \figref{Identifications_RingDiagrams_examples}a.
When one considers the two pairs of nodes connected by dottes lines
as one node, it can be seen as three concatenated rings of length
$2$, so a pseudo-cactus diagram. Complementing it with its counter
diagrams, we can convince ourselves that it cancels (with the prefactors
$\frac{1}{4}-\frac{1}{2}+\frac{1}{4}$). We now proceed to the diagram
of eighth order depicted in \figref{Identifications_RingDiagrams_examples}b,
in which we identify the indices of one pair of nodes opposite in
the ring. There are four of these pairs to choose from, so that the
prefactor is $\frac{1}{16}\cdot4=\frac{1}{4}$. We can as well build
this diagram with one of the half-rings replaced by a $\frac{\partial\GamJ_{V}}{\partial v}$-vertex,
yielding the prefactor $-\frac{1}{2}$ or both replaced, yielding
$\frac{1}{4}$. This contribution therefore cancels as well. We next
choose not only one pair of nodes to bear to same index, but two of
them, the second pair ``rotated'' by 90 degrees with respect to
the first one, as in \figref{Identifications_RingDiagrams_examples}c.
We will call this a crossing identification (compare the notion of
crossing partitions in the study of free probability theory \cite{Pappalardi22_170603}).
We here have only $2$ possibilities to pick these two pairs and therefore
the prefactor for the diagram composed only of $K$s is $\frac{1}{16}\cdot2=\frac{1}{8}$.
The counter diagrams, however, keep their prefactors because we fix
one pair of indices to cut the ring and replace the sub-diagrams by
$\frac{\partial\GamJ_{V}}{\partial v}$ and the second pair of indices
is already fixed by this choice. Therefore, if there are multiple
pairs of nodes with indices identified in such a way that one cannot
represent the resulting diagram as a pseudo-cactus there is no guarantee
that this contribution is canceled. This is due to the fact that in
these cases  there are not necessarily  matching counter diagrams
for all pairs of nodes with identical indices. Put differently, the
counter diagrams fail to cancel the original contributions because
the diagram with the crossing identification has a different (higher)
symmetry than the diagram with the single pair of nodes. This is analogous
to the case of the contributions from the ring diagrams in fourth
order in $K$ with one or two pairs of nodes with identical indices,
compare \eqref{Ring_diagrams_fourth_order_GamJ_one_pair_equal}, \eqref{Ring_diagrams_fourth_order_GamJ_two_pairs_equal}
and their discussion. 

We note for completeness, however, that there are also cases of diagrams
with crossing identifications that do cancel, like in \figref{Identifications_RingDiagrams_examples}e.
There, as well as in \figref{Identifications_RingDiagrams_examples}d,
there are $3$ possibilities to choose the respective pairs from.
Because the symmetry of \figref{Identifications_RingDiagrams_examples}d
is decisive for the prefactor of the counter diagrams, also the diagram
with crossing identification gets canceled here - as the number of
possible identifications of pairs is the same in both settings. To
study these cases more generally, one would need to systematically
count all possibilities to choose the pairs of nodes contributing
to the crossing identifications. This can be achieved applying group
theory to the symmetry group corresponding to this choice - related
approaches are pursued in the field of enumerative combinatorics (see,
e.g., \cite[chap. 7]{Cameron17_book}). This could potentially allow
us to figure classes of crossing identifications with the same symmetry
factor as the non-crossing identifications and therefore further cancellations
- we leave an investigation of this question for future work.

Either way, we have now seen that pseudo-cactus diagrams without crossing
identifications, like that in \figref{Identifications_RingDiagrams_examples}a,
get canceled. For proper cactus diagrams, these complications do not
exist because one does not have to identify possible pseudo nodes,
the cumulants are simply what they are.

Summarized, the following types of diagrams always cancel:
\begin{enumerate}
\item Cactus diagrams
\item Pseudo-cactus diagrams without crossing identifications
\end{enumerate}
\selectlanguage{american}%
In addition to 2., all pseudo-cactus diagrams with crossing identifications
cancel in which the number possibilities to pick the pairs of nodes
for identification of indices equals the number of possibilities to
pick just one pair of the crossing identification. We keep this point
separate because we don't need it for our application. 

\selectlanguage{english}%
Before closing this section, we note that we did not characterize
all diagrams contributing to $\GamJ$, but only those of the same
type as those contributing to $W$. However, $\GamJ$ includes of
course contributions with cumulants of orders higher than $2$. Differentiating
those with respect to $m_{i}$ or $v_{i}$ requires to take the derivative
of the cumulant with respect to $j_{i}$ and $k_{i}$ to obtain the
dependence on $m_{i}$ and $v_{i}$ via the inner derivatives. The
latter is basically given by the Hessian of $\GamJ_{0}$, which for
completeness we derive in \subsecref{Hessian_GamJ}. These $\GamJ$-derivatives
serve as vertices in the construction of higher orders - the lowest
one being $6$. However, in the application we will address in \subsecref{Application_pSpins},
we do not need them because their contribution is sub-leadling in
the thermodynamic limit.

\subsection{The Gaussian Legendre transform: $\protect\Gamc$\label{subsec:Gaussian_LegTrafo}}

In addition to the Legendre transforms on the one-point source fields
$\j$ and $\k$, one can also transform with respect with respect
to the two-point source field $K$, therefore fixing the covariances,
which we denote as
\[
\fl\Gamc\left[\boldsymbol{\m},\boldsymbol{\v},c\right]:=\sup_{\boldsymbol{j},\boldsymbol{k},K}\left[\sum_{i}\left(\m_{i}j_{i}+\left(\v_{i}+m_{i}^{2}\right)k_{i}\right)+\frac{1}{2}\text{tr}\left(K\left(c+\boldsymbol{\m}\boldsymbol{\m}^{\T}\right)\right)-W\left[\boldsymbol{j},\boldsymbol{k},K\right]\right].
\]
We will call this the Gaussian Legendre transform because it depends
on all cumulants present in a Gaussian theory. When applied to an
actual Gaussian theory it is therefore what Vasiliev calls a complete
Legendre transform \cite[sec. 6.2.]{vasiliev2019functional}. Its
equations of state are
\begin{eqnarray*}
\fl\frac{\partial\Gamc}{\partial m_{i}} & = & j_{i}+2m_{i}k_{i}+\sum_{j\left(\neq i\right)}K_{ij}m_{j}\\
\fl\frac{\partial\Gamc}{\partial v_{i}} & = & k_{i}\\
\fl\frac{\partial\Gamc}{\partial c_{ij}} & = & K_{ij}.
\end{eqnarray*}

\subsubsection{Derivation of a perturbation expansion for $\protect\Gamc$\label{subsec:Derivation_diagrams_GammaC}}

To derive the iteration equation for a diagrammatic expansion in the
covariance $c$ for this quantity, we proceed analogously to \subsecref{Derivation_expansion_GamJ}
to obtain 
\begin{equation}
\fl\exp\left(-\Gamc_{v}\left[\boldsymbol{\m},\boldsymbol{\v}\right]\right)=e^{-W_{0}\left[\boldsymbol{j}_{0},\boldsymbol{k}_{0}\right]}\underset{L\rightarrow\infty}{\lim}\left(1+\frac{1}{L}A_{c}\left(\frac{\partial}{\partial\boldsymbol{j}}\right)\right)^{L}\left.e^{W_{0}\left[\boldsymbol{j},\boldsymbol{k}\right]}\right|_{\boldsymbol{j}=\boldsymbol{j}_{0},\,\boldsymbol{k}=\boldsymbol{k}_{0}},\label{eq:Gamc_v_exp_L_representation}
\end{equation}
where

\begin{eqnarray}
\fl A_{c}\left(\frac{\partial}{\partial\boldsymbol{j}}\right) & := & \left(\frac{\partial\Gamc_{V}}{\partial\boldsymbol{m}}\right)^{\T}\left(\frac{\partial}{\partial\boldsymbol{j}}-\boldsymbol{m}\right)\label{eq:Ac_operator_del_m}\\
\fl &  & +\left(\frac{\partial\Gamc_{V}}{\partial\boldsymbol{v}}\right)^{\T}\left[\left(\frac{\partial}{\partial\boldsymbol{j}}-\boldsymbol{\m}\right)^{2}-\boldsymbol{v}\right]\label{eq:Ac_operator_del_v}\\
\fl &  & +\frac{1}{2}\mathrm{tr}\left(\frac{\partial\Gamc_{V}}{\partial c}\left[\left(\frac{\partial}{\partial\boldsymbol{j}}-\boldsymbol{\m}\right)\left(\frac{\partial}{\partial\boldsymbol{j}}-\boldsymbol{\m}\right)^{\T}-c\right]\right).\label{eq:Ac_operator_del_c}
\end{eqnarray}

Note that because the interaction has the form of a an off-diagonal
two-point source field, which we have replaced by $\frac{\partial\Gamc_{V}}{\partial c}$,
$H_{V}$ does not occur here anymore and also that the matrix product
inside the $\mathrm{tr}$-operator runs only over the off-diagonal
entries because $\frac{\partial\Gamc_{V}}{\partial c}$ has only off-diagonal
entries by definition. The tricky part now is that for $\Gamc$, the
expansion parameter is $c$ instead of $K$. Therefore, $\frac{\partial\Gamc_{V}}{\partial c}$
is of one order lower than $\Gamc$. Consequently, evaluating \eqref{Gamc_v_exp_L_representation}
at a fixed order $n$ in $c$ involves $c$-derivatives of the $n$-th
order term also on the right-hand side - a priori. However, we will
see that these terms are canceled for $n>2$ (compare also \cite[sec. 5.1.4.]{Kuehn23_115001}).
In order to see this, we treat the order $2$ separately. Here, it
is straightforward to perform the additional Legendre transform with
respect to $K$ explicitly, obtaining $\Gamc$ from $\GamJ$:
\begin{eqnarray}
\fl\Gamc\left[\boldsymbol{\m},\boldsymbol{\v},c\right] & = & \sup_{K}\left[\frac{1}{2}\text{tr}\left(K\left(c+\boldsymbol{\m}\boldsymbol{\m}^{\T}\right)\right)+\GamJ\left[\boldsymbol{\m},\boldsymbol{\v},K\right]\right],\label{eq:Gamc_from_GamK}\\
\fl & = & \sup_{K}\left[\frac{1}{2}\sum_{i\neq j}K_{ij}\left(c_{ij}+\m_{i}\m_{j}\right)+\GamJ\left[\boldsymbol{\m},\boldsymbol{\v},K\right]\right]\nonumber 
\end{eqnarray}
where
\[
\fl\GamJ\left[\boldsymbol{\m},\boldsymbol{\v},K\right]=\GamJ_{0}\left[\boldsymbol{\m},\boldsymbol{\v}\right]-\frac{1}{2}\sum_{i\neq j}K_{ij}m_{i}m_{j}-\frac{1}{4}\sum_{i\neq j}K_{ij}^{2}v_{i}v_{j}+\mathcal{O}\left(K^{3}\right).
\]
Evaluating the condition on $K$ given by the supremum by differentiating
with respect to $K_{ij}$, we obtain
\[
\fl c_{ij}=K_{ij}v_{i}v_{j}.
\]
Solving for $K_{ij}$ and inserting into \eqref{Gamc_from_GamK},
we obtain\footnote{This derivation demonstrates that \cite[eq. (40)]{Kuehn23_115001}
is actually the correct second-order contribution to the free energy
at fixed covariance and not just a consequence of a specific solution
to the tensor equation \cite[eq. (38)]{Kuehn23_115001}.}
\begin{equation}
\fl\Gamc\left[\boldsymbol{\m},\boldsymbol{\v},c\right]=\GamJ_{0}\left[\boldsymbol{\m},\boldsymbol{\v}\right]+\frac{1}{4}\sum_{i\neq j}\frac{c_{ij}^{2}}{v_{i}v_{j}}+\mathcal{O}\left(c^{3}\right).\label{eq:Gamma_c_second_order}
\end{equation}
Considering higher orders now, we obtain as our first estimate of
$\frac{\partial\Gamc_{V}}{\partial c}$ entering into \eqref{Ac_operator_del_c}
\begin{equation}
\fl\frac{\partial\Gamc_{V}}{\partial c_{ij}}=\frac{c_{ij}}{v_{i}v_{j}}+\mathcal{O}\left(c^{2}\right).\label{eq:Deriv_v_Gamma_c_first_order}
\end{equation}
Now we can address the issue of what happens with the $c$-derivative
of the terms of $n$-th order on the right-hand side of \eqref{Gamc_v_exp_L_representation},
evaluated at order $n>2$: the contribution from $\frac{\partial\Gamc_{V}}{\partial c}$
has to be combined with a term of order $1$ to yield a $n$-th-order
term, so either the $-c$ in the definition of $A_{c}$, \eqref{Ac_operator_del_c},
or the derivative of the second-order contribution to $\Gamc$ with
respect to $c$. With \eqref{Deriv_v_Gamma_c_first_order}, we see
that both resulting terms are actually equal, up to the sign, and
therefore cancel. For a more detailed discussion of this point, see
\cite[sec. 5.1.4.]{Kuehn23_115001}. 

Therefore, from this point on, we can proceed largely analogous to
the construction of the diagrammatic expansion of $\GamJ$, the only
difference being that now, the newly generated vertices are not only
coming from derivatives of the single-spin cumulants, represented
by nodes, but also the covariances, which the edges represent now
- slightly abusing diagrammatic notation, as in

\begin{fmffile}{Redefinition_edges}	
	\begin{eqnarray}
		\fl \mkern 10mu
		\parbox{25mm}{
			\begin{fmfgraph*}(25,25)
				\fmfpen{0.5thin}
				\fmftop{o1,o2,o3}
				\fmfbottom{u1,u2,u3}
				\fmf{plain}{u1,o2}
				\fmf{plain}{u3,o2}
				\fmfv{label=$i$, label.angle=-90, label.dist=4.5pt}{u1}
				\fmfv{label=$j$, label.angle=-90, label.dist=4.5pt}{u3}
			\end{fmfgraph*}
		} \mkern-60mu = \frac{c_{ij}}{v_{i}v_{j}},\\
		\nonumber
	\end{eqnarray}
\end{fmffile}whereas nodes still represent cumulants of the uncoupled theory. For
a detailed discussion of the contributions to this expansion coming
about by the terms also present in the differential operator $A_{c}$
without transform with respect to $\boldsymbol{k}$ - \eqref{Ac_operator_del_m}
and \eqref{Ac_operator_del_c} - up to order four, we refer to \cite[sec. 5.1.5. and 5.1.6.]{Kuehn23_115001}
and we will limit ourselves to studying the differences occurring
due to the new part of $A_{c}$, \eqref{Ac_operator_del_v}, which
starts contributing at fourth order. 

\subsubsection{The fourth order of the Gaussian Legendre transform}

In fourth order, \eqref{Ac_operator_del_v} generates two contributions,
largely analogous to \eqref{A_operator_del_v}. Note however, that
here, the dependence on $v_{i}$ is a different one than for $\GamJ$
because it is contained not only through the single-spin contributions,
but also in the interaction. However, differentiating the second-order
correction in \eqref{Gamma_c_second_order} with respect to $v_{i}$,
we obtain

\begin{fmffile}{Deriv_Gamma_c_v_scd_order}
	\begin{eqnarray}
		\fl\frac{\partial}{\partial v_{k}}\frac{1}{4}\sum_{i\neq j}\frac{c_{ij}^{2}}{v_{i}v_{j}}=-\frac{1}{2}\sum_{i\left(\neq k\right)}\frac{c_{ik}^{2}}{v_{i}v_{k}^{2}}=-\frac{1}{2}\sum_{i\left(\neq k\right)}\left(\frac{c_{ik}}{v_{i}v_{k}}\right)^{2}v_{i}
		=-\mkern 20mu\parbox{20mm}{
			\begin{fmfgraph*}(20,30) 				
				\fmfpen{0.5thin} 				
				\fmftop{o1,o2} 				
				\fmfbottom{u1,u2}
				\fmf{plain}{o1,v1,u1}
				\fmf{plain}{o1,o2}
				\fmf{plain}{u1,u2}
				\fmf{dashes}{o2,u2}
				\fmfv{decor.shape=circle,decor.filled=full, decor.size=6.5thin}{v1}
				\fmfv{label=$k$, label.angle=0, label.dist=4.5pt}{o2}
				\fmfv{label=$k$, label.angle=0, label.dist=4.5pt}{u2}	 			
			\end{fmfgraph*} 		
		}\mkern-36mu,
	\end{eqnarray}
\end{fmffile}so the analogous contribution to that of $\GamJ$. Consequently, it
has the same effect on the spectacles diagram as for $\GamJ$: it
gets canceled, as in \eqref{Glasses_diagrams_fourth_order_GamJ}.

Then, \eqref{Ac_operator_del_v} generates a ring diagram with two
pairs of indices identified, with the prefactor $-\frac{1}{4}$, when
both halves of the ring come from $\frac{\partial\Gamc_{V}}{\partial\boldsymbol{v}}$,
and $\frac{1}{2}$ when only one does. This contribution is combined
with the other terms, generated by \eqref{Ac_operator_del_c}, therefore
involving $\frac{\partial\Gamc_{V}}{\partial c}$. 

The reasoning to determine the symmetry factors for the remaining
terms is very similar to the contributions described before and it
has been presented in \cite[eq. (58) - (61)]{Kuehn23_115001}. We
therefore do not discuss it in detail here and simply state the result
that by taking all terms together, we obtain

\begin{fmffile}{Ring_diagrams_fourth_order_Gamc} 	
	\begin{eqnarray}
	 \fl & 
	- \mkern 20mu
		\parbox{25mm}{
			\begin{fmfgraph*}(40,30) 				
				\fmfpen{0.5thin} 				
				\fmftop{o1,o2,o3} 				
				\fmfbottom{u1,u2,u3}
				\fmf{plain}{o1,v1,u1}
				\fmfleft{l1}
				\fmfright{r1}
				\fmf{plain}{o3,v2,u3}
				\fmf{plain}{o1,o2,o3}
				\fmf{plain}{u1,u2,u3}
				\fmfv{decor.shape=circle,decor.filled=empty, decor.size=6.5thin}{l1,r1,o2,u2}				 			
			\end{fmfgraph*} 		
		}
		\mkern-30mu + \mkern 45mu
		\parbox{25mm}{
			\begin{fmfgraph*}(40,30) 				
				\fmfpen{0.5thin} 				
				\fmftop{o1,o2,o3} 				
				\fmfbottom{u1,u2,u3}
				\fmf{plain}{o1,v1,u1}
				\fmfleft{l1}
				\fmfright{r1}
				\fmf{plain}{o3,v2,u3}
				\fmf{plain}{o1,o2,o3}
				\fmf{plain}{u1,u2,u3}
				\fmf{dots, tension = 100}{o2,u2}
				\fmfv{decor.shape=circle,decor.filled=empty, decor.size=6.5thin}{r1,o2,u2}
				\fmfv{decor.shape=circle,decor.filled=full, decor.size=6.5thin}{l1}			 			
				\fmfv{label=$\frac{\partial \Gamc_{V}}{\partial \boldsymbol{c}}$, label.angle=180, label.dist=4.5pt}{l1}
			\end{fmfgraph*} 		
		}
		\mkern-30mu - \mkern 45mu
		\parbox{25mm}{
			\begin{fmfgraph*}(40,30) 				
				\fmfpen{0.5thin} 				
				\fmftop{o1,o2,o3} 				
				\fmfbottom{u1,u2,u3}
				\fmfleft{l1}
				\fmfright{r1}
				\fmf{plain}{o1,l1,u1}
				\fmf{plain}{o3,r1,u3}
				\fmf{plain}{o1,o2,o3}
				\fmf{plain}{u1,u2,u3}
				\fmf{dots, tension = 100}{o2,u2}
				\fmfv{decor.shape=circle,decor.filled=empty, decor.size=6.5thin}{o2,u2}
				\fmfv{decor.shape=circle,decor.filled=full, decor.size=6.5thin}{l1,r1}
				\fmfv{label=$\frac{\partial \Gamc_{V}}{\partial \boldsymbol{c}}$, label.angle=180, label.dist=4.5pt}{l1}
				\fmfv{label=$\frac{\partial \Gamc_{V}}{\partial \boldsymbol{c}}$, label.angle=0, label.dist=4.5pt}{r1}
			\end{fmfgraph*} 		
		}\label{eq:Ring_fourth_order_simple_and_unequal}\\[5mm]
		 \fl & 
		\mkern 40mu + \mkern 55mu
		\parbox{25mm}{
			\begin{fmfgraph*}(40,30) 				
				\fmfpen{0.5thin} 				
				\fmftop{o1,o2,o3} 				
				\fmfbottom{u1,u2,u3}
				\fmf{plain}{o1,v1,u1}
				\fmfleft{l1}
				\fmfright{r1}
				\fmf{plain}{o3,v2,u3}
				\fmf{plain}{o1,o2,o3}
				\fmf{plain}{u1,u2,u3}
				\fmf{dashes, tension = 100}{o2,u2}
				\fmfv{decor.shape=circle,decor.filled=empty, decor.size=6.5thin}{r1,o2,u2}
				\fmfv{decor.shape=circle,decor.filled=full, decor.size=6.5thin}{l1}
				\fmfv{label=$\frac{\partial \Gamc_{V}}{\partial \boldsymbol{v}}$, label.angle=180, label.dist=4.5pt}{l1}
			\end{fmfgraph*} 		
		}
		\mkern-30mu - \mkern 45mu
		\parbox{25mm}{
			\begin{fmfgraph*}(40,30) 				
				\fmfpen{0.5thin} 				
				\fmftop{o1,o2,o3} 				
				\fmfbottom{u1,u2,u3}
				\fmfleft{l1}
				\fmfright{r1}
				\fmf{plain}{o1,l1,u1}
				\fmf{plain}{o3,r1,u3}
				\fmf{plain}{o1,o2,o3}
				\fmf{plain}{u1,u2,u3}
				\fmf{dashes, tension = 100}{o2,u2}
				\fmfv{decor.shape=circle,decor.filled=empty, decor.size=6.5thin}{o2,u2}
				\fmfv{decor.shape=circle,decor.filled=full, decor.size=6.5thin}{l1,r1}
				\fmfv{label=$\frac{\partial \Gamc_{V}}{\partial \boldsymbol{v}}$, label.angle=180, label.dist=4.5pt}{l1}
				\fmfv{label=$\frac{\partial \Gamc_{V}}{\partial \boldsymbol{v}}$, label.angle=0, label.dist=4.5pt}{r1}
			\end{fmfgraph*} 		
		}\mkern -10mu,
	\end{eqnarray} 
\end{fmffile}where again dotted lines denote pairs of nodes with unequal indices.
This translates into
\begin{eqnarray}
\fl &  & -\frac{1}{8}\sum_{i\neq j\neq k\neq l\neq i}K_{ik}\v_{k}K_{kj}\v_{j}K_{jl}\v_{j}K_{li}\v_{i}\label{eq:Translation_pure_ring_fourth_order}\\
\fl &  & +\frac{1}{2}\sum_{i\neq j}\sum_{k,l\left(\neq i,j\right)}K_{ik}\v_{k}K_{kj}\v_{j}K_{jl}\v_{l}K_{li}\v_{i}-\frac{1}{4}\sum_{i\neq j}\sum_{k,l\left(\neq,j\right)}K_{ik}\v_{k}K_{kj}\v_{j}K_{jl}\v_{l}K_{li}\v_{i}\\
\fl &  & +\frac{1}{2}\sum_{i\neq j}\sum_{k\left(\neq i,j\right)}\v_{i}K_{ik}^{2}\v_{k}^{2}K_{jk}^{2}\v_{j}-\frac{1}{4}\sum_{i\neq j}\sum_{k\left(\neq i,j\right)}\v_{i}K_{ik}^{2}\v_{k}^{2}K_{jk}^{2}\v_{j}\\
\fl & = & \frac{1}{8}\sum_{\overset{i\neq j\neq k\neq l\neq i,}{i\neq k,j\neq l}}K_{ik}\v_{k}K_{kj}\v_{j}K_{jl}\v_{j}K_{li}\v_{i}+\frac{1}{4}\sum_{i\neq j}\sum_{k\left(\neq i,j\right)}\v_{i}K_{ik}^{2}\v_{k}^{2}K_{jk}^{2}\v_{j}+\frac{1}{8}\sum_{i\neq j}\v_{i}\v_{i}^{4}K_{ij}^{4}\v_{j}^{4},\nonumber 
\end{eqnarray}
where we write $b\left(\neq a\right)$ in sums to indicate that we
sum over the index $b$, restricting it not to equal $a$, over which
we do not sum. This result can be expressed much shorter as
\[
\fl\frac{1}{8}\sum_{i\neq j\neq k\neq l\neq i}K_{ik}\v_{k}K_{kj}\v_{j}K_{jl}\v_{j}K_{li}\v_{i}=\frac{1}{8}\mathrm{tr}\left(\left(Kv\right)^{4}\right).
\]
In other words: fixing also the (on-site) variance, in addition to
the covariances, leads to a simplification because we do not have
to differentiate between subsets of indices with different numbers
of them identified. Indeed, without the additional part of the differential
operator $A_{c}$ in \eqref{Ac_operator_del_v}, we would only have
the diagrams in \eqref{Ring_fourth_order_simple_and_unequal}, the
latter two of which with the restriction that one pair of opposite
nodes bears unequal indices. Only by the other two diagrams, this
restriction is lifted because it adds exactly the complementary contributions.
This is actually a general rule, as we will discuss in the next section.

\subsubsection{Ring diagrams in the Gaussian Legendre transform\label{subsec:Gauss_LegTrafo_Loops}}

We can generalize the insights gained for the fourth-order ring diagram
to arbitrary order, which will allow us to resum this class of diagrams.

In \cite{Kuehn23_115001}, the sub-class of ring diagrams with indices
that are pairwise unequal was characterized. If the index set is continuous
(like, e.g., for simple liquids), this is sufficient for a straight-forward
resummation. Otherwise it is not because the condition on all indices
to be pairwise unequal prevents expressing this problem as a simple
matrix equation. However, this is different if one includes an additional
transform with respect to the variances. In this case one obtains
additional contributions according to \eqref{Ac_operator_del_v},
which fixes this problem. 

We will demonstrate that the prefactor of the ring diagram of order
$n$ is $\frac{\left(-1\right)^{n}}{2n}$ - and, in contrast to \cite{Kuehn23_115001},
the sum over the multiple indices it implies bears no further restriction
than that that directly neighbored nodes bear unequal indices. When
constructing a ring diagram of order $n$, the following components
are provided:
\begin{itemize}
\item concatenations of interactions, which we will call snake diagrams,
of length $k=1$,$2$,...,$n-2$ with the condition that the outer
indices are unequal, coming about from \eqref{Ac_operator_del_c},
with symmetry factor $\left(-1\right)^{k+1}$
\item snake diagrams of length $2$,...,$n-2$ with the condition that the
outer indices are equal, coming about from part \eqref{Ac_operator_del_v}
of the operator $A$, with symmetry factor $\frac{1}{2}\left(-1\right)^{k+1}$
\end{itemize}
The sign of both contribution derives from the respective ring diagrams
the snakes come from: the first comes about by differentiating the
ring diagram of order $k+1$ with respect to the covariance, hence
$\left(-1\right)^{k+1}$, the second comes from differentiating the
ring diagram of order $k$ (sign $\left(-1\right)^{k}$) with respect
to a variance. In the latter case, the dependence is of the form $\frac{1}{v_{i}}$,
hence the prefactor of this snake diagram is $\left(-1\right)^{k+1}$
as well. The difference in the symmetry factor comes about by the
fact that in the first case, one differentiates with respect to a
two-point quantity (the covariance), whereas in the second case, one
differentiates with respect to a one-point quantity (the variance).
Inserted into \eqref{Ac_operator_del_c} and \eqref{Ac_operator_del_v},
respectively, one observes that both contributions act in exactly
the same way on the unperturbed cumulants, only with opposite - or,
rather, complementary - restrictions on the outer indices, because
\eqref{Ac_operator_del_c} has an additional factor $\frac{1}{2}$
compared to \eqref{Ac_operator_del_v}. Because of the many ways one
can partition $n$ to subdivide the ring into snakes\footnote{For being precise, one has to add: discarding the option to include
snake diagrams of order $n-1$, according to the rules derived in
\cite{Kuehn23_115001}.} it is far from obvious that the resulting prefactor is indeed $\frac{\left(-1\right)^{n}}{2n}$.
However, this result is proven in \cite[sec. 8 and Appendix]{Kuehn23_115001},
without alluding to the fact that indices are pairwise unequal, actually
not even referring to the summation at all, but assuming that the
sums are performed for the right index set. With the additional contribution
of two-point interactions with identical interactions, we have lifted
the need to distinguish these cases because no matter how the ring
is partitioned, there are no additional restrictions imposed on the
indices.

Therefore, the argument of \cite{Kuehn23_115001} for the resummation
of ring diagrams now literally carries over, only without the restriction
to pairwise unequal indices. This has to be so because for a Gaussian
theory, one knows that the entropy amounts to
\begin{eqnarray}
\fl S-\frac{N}{2}\ln\left(2\pi e\right)=\frac{1}{2}\ln\left(\det\left(c\right)\right) & = & \frac{1}{2}\sum_{i}\ln\left(V_{i}\right)+\frac{1}{2}\mathrm{tr}\left(\sum_{n=2}^{\infty}\frac{\left(-1\right)^{n+1}}{n}\left(\frac{c_{\neq}}{\sqrt{V}\sqrt{V}}\right)^{n}\right)\label{eq:Entropy_Gauss}\\
\fl\mathrm{where}\ \left(\frac{c_{\neq}}{\sqrt{V}\sqrt{V}}\right)^{n} & = & \sum_{i_{1}\neq i_{2}\neq\dots\neq i_{n-1}\neq i_{1}}\frac{c_{i_{1}i_{2}}}{V_{i_{1}}}\frac{c_{i_{2}i_{3}}}{V_{i_{2}}}\dots\frac{c_{i_{n-1}i_{1}}}{V_{i_{n-1}}}.
\end{eqnarray}
In the sum into which the matrix power translates, the only restriction
is upon neighbored indices. This corresponds to the resummation of
all ring diagrams without every single pair of indices having to be
pairwise unequal. 

\section{Applications}

\subsection{Spins coupled by rotationally invariant matrices\label{subsec:Application_pSpins}}

We here consider the example from Maillard et al. \cite{Maillard19_113301}
to study message-passing algorithms. They consider an interacting
part of the Hamiltonian of the form
\[
H_{V}=-\frac{1}{2}\sum_{1\leq i\neq j\leq N}x_{i}K_{ij}x_{j},
\]
where $J$ is a symmetric rotationally invariant matrix (their model
S) and the $x_{i}$ are spherical spins (the type of $x_{i}$ does
not matter for us, but choosing it as a spherical spin has the merit
that it allows them fto derive the exact expression of the free energy
in the thermodynamic limit by other means than a small-coupling expansion.
There is therefore a ground truth to compare with). We have excluded
diagonal terms, unlike Maillard et al. do, however, we can absorb
it into the two-point source-field, so both conventions are fine.
Extending the work of Plefka and Georges/Yedidia \cite{plefka1982convergence,georges1991expand}
to the Legendre transform at fixed mean and variance, Maillard et
al. set up a small-coupling expansion. It is, however, not diagrammatic;
the diagrams they are drawing are depictions of the terms they have
derived by other means. In addition, they build on a result from random-matrix
theory by Guionnet et al. \cite{Guionnet05_435} to show that out
of these contributions only those remain that they call simple cycles
and cacti, all other vanish in the thermodynamic limit of $N\rightarrow\infty$.
Their simple cycle (see their sec. 5.2) corresponds to a ring diagram
with all indices pairwise unequal in our terminology. Cactus diagrams
in their terminology comprise what we call cactus and pseudo-cactus
diagrams with all sub-diagrams being simple cycles (because the other
ones are irrelevant in the thermodynamic limit). They convince themselves
by explicit computation that up to fourth order that the cactus diagrams
are canceled in the small-coupling expansion. They therefore conjecture
the free energy to be (their eq. (25), note that they scale their
free energy with $\frac{1}{N}$, which we do not)

\begin{equation}
\fl\GamJ=\GamJ_{0}+\frac{1}{2}\sum_{i\neq j}K_{ij}m_{i}m_{j}+\sum_{p=2}^{\infty}\frac{1}{2p}\sum_{\overset{i_{1},\dots,i_{p}}{\mathrm{pairwise\ }\mathrm{distinct}}}v_{i_{1}}K_{i_{1}i_{2}}v_{i_{2}}K_{i_{2}i_{3}}\dots K_{i_{1}i_{p}}+o_{N}\left(N\right),\label{eq:Interaction_correction_Maillard}
\end{equation}
assuming that cactus diagrams cancel in all orders. However, the formalism
of Georges and Yedidia \cite{georges1991expand} ``can not (somehow
disappointingly) give an analytic result for an arbitrary perturbation
order $n$'', as Maillard et al. write.

With our insights from \subsecref{Cancellation_higher_orders}, we
now fill this gap and complete their proof, rendering their conjecture
into a theorem. The only qualm that one could have is that, as discussed
in \subsecref{Cancellation_higher_orders}, we actually do not cancel
all pseudo-cactus diagrams. However, the ones that are not canceled
are the ones with the crossing identities and they are therefore of
the type that Maillard et al. call strongly irreducible - and therefore
vanish in the thermodynamic limit. Indeed, the only ``dangerous''
in the sense of Maillard et al. are cacti (cacti and pseudo cacti
in our sense) composed of simples cycles - and for these we construct
a counter diagram with matching prefactor in all of the cases - due
to the very structure of our diagrammatics.

\subsection{The Gaussian Legendre transform as a (maximum) entropy\label{subsec:Application_MaxEnt}}

An important application of our results on the joint Legendre transform
with respect to $\j$, $\k$ and $K$ is to estimate the maximum value
for the entropy given the first two moments of some data set. The
corresponding probability distribution is indeed given by \cite{Jaynes03}
\[
\fl P_{\mathrm{MaxEnt}}\left(\boldsymbol{n}\right)=\frac{1}{{\cal Z}\left(\j,\k,K\right)}e^{\sum_{i}\left(j_{i}n_{i}+k_{i}n_{i}^{2}\right)+\sum_{i\neq j}K_{ij}n_{i}n_{j}}
\]
and the corresponding entropy therefore is
\[
\fl S=\left\langle -\sum_{i}\left(j_{i}n_{i}+k_{i}n_{i}^{2}\right)-\sum_{i\neq j}K_{ij}n_{i}n_{j}\right\rangle _{\boldsymbol{n}}+\ln\left({\cal Z}\left(\j,\k,K\right)\right).
\]
We can express the fact that the source fields are chosen such that
the first two moments of the model fit the measured ones by adding
a supremum, yielding
\begin{equation}
\fl S=\text{\ensuremath{\sup_{\j,\k,K}\left[\left\langle -\sum_{i}\left(j_{i}n_{i}+k_{i}n_{i}^{2}\right)-\sum_{i\neq j}K_{ij}n_{i}n_{j}\right\rangle _{\boldsymbol{n}}+\ln\left({\cal Z}\left(\j,\k,K\right)\right)\right]}=-\ensuremath{\Gamc\left[\boldsymbol{\m},\boldsymbol{v},c\right]}.}\label{eq:Identity_Gamc_Entropy}
\end{equation}
With the relation \eqref{Identity_Gamc_Entropy} at hand, we can use
the results from \subsecref{Gaussian_LegTrafo}, to estimate entropies
from empirical probability distributions. This is particularly useful
when the data is poorly sampled and a direct estimate of the entropy
is therefore prone to biases \cite{Paninski03_1191}. In particular,
\subsecref{Gauss_LegTrafo_Loops} then enables a stable estimate,
without being too restrictive with respect to the form of the underlying
probability distribution. In particular, it avoids assuming a Gaussian
shape, but allows us to exactly take into account the single-spin
contributions to the entropy.

We can also imagine to not only fix the first two, but the whole single-spin
probability distribution or all of its cumulants (which is equivalent
given that all cumulants exist). This would add another term to \eqref{Ac_operator_del_m}
for every additional condition. 

For every finite order, of course, only a finite number of this terms
contribute - in particular, the term in the differential operator
$A_{c}$ that comes about by fixing the $n$-th-order cumulant is
of order $n$ in the covariance (coming from the diagram with just
two cumulants, connected by $n$ interactions). Therefore the lowest-order
correction to the free energy occurring because the variance is fixed,
\eqref{Ac_operator_del_v}, is of order four. In any case, if one
only considers diagrams of the ring type, the additional contributions
to $A_{c}$ are insubstantial because they generate diagrams of different
topologies by construction. While the resummation of rings diagrams
is usually an ad-hoc approach, there are cases in which it can we
justified: if the covariances obey the mean-field scaling $1/N$ with
$N$ the number of spins for example (and higher-order correlations
negligible), the ring diagrams are the only remaining ones for $N\rightarrow\infty$.

\subsection{Ising model\label{subsec:Application_Ising}}

The Ising model is special, because due to its binary character, one
source field fixes all single-unit moments. This means that Legendre
transforms of order higher than one are not well-defined for all arguments.
However, there is no problem if we evaluate the free energy only at
the values given by the Ising model, that is
\begin{eqnarray*}
\fl\left\langle n^{2k+1}\right\rangle  & = & m\\
\fl\left\langle n^{2k}\right\rangle  & = & 1.
\end{eqnarray*}

\subsubsection{Fourth-order small-coupling expansion}

It is not only possible, but also useful to formally define higher-order
Legendre transforms because, as we will see, it simplifies the diagrammatics
and explains puzzles that would remain unresolved if only the first
transform were considered.

Actually, the first contributions to $\GamJ$ in addition to that
of the first Legendre transform, coming about from \eqref{A_operator_del_v},
are those discussed in \subsecref{Cancellation_fourth_order_GamJ}.
They contribute to the fourth order, which in total reads 

\begin{fmffile}{Fourth_order_complete_with_compensation}	
	\begin{eqnarray}
		\fl -\parbox{25mm}{
			\begin{fmfgraph*}(75,25)
				\fmfpen{0.5thin}
				\fmftop{o1,o2,o3}
				\fmfbottom{u1,u2,u3}
				\fmf{phantom}{u1,dul,dml,v1,o2}
				\fmf{plain}{dml,v1,o2}
				\fmf{phantom}{u2,v2,dml,dol,o1}
				\fmf{plain}{u2,v2,dml}
				\fmf{phantom}{u2,v3,dmr,dor,o3}
				\fmf{plain}{u2,v3,dmr}
				\fmf{phantom}{u3,dur,dmr,v4,o2}
				\fmf{plain}{dmr,v4,o2}
				\fmfv{decor.shape=circle,decor.filled=empty, decor.size=6.5thin}{v1,v2,v3,v4}
			\end{fmfgraph*}
		}
		+\mkern-10mu \parbox{25mm}{
			\begin{fmfgraph*}(75,25)
				\fmfpen{0.5thin}
				\fmftop{o1,o2,o3}
				\fmfbottom{u1,u2,u3}
				\fmf{phantom}{u1,dul,dml,v1,o2}
				\fmf{plain}{dml,v1,o2}
				\fmf{phantom}{u2,v2,dml,dol,o1}
				\fmf{plain}{u2,v2,dml}
				\fmf{phantom}{u2,v3,dmr,dor,o3}
				\fmf{plain}{u2,v3,dmr}
				\fmf{phantom}{u3,dur,dmr,v4,o2}
				\fmf{plain}{dmr,v4,o2}
				\fmf{dashes, tension=0}{v2,v4}
				\fmfv{decor.shape=circle,decor.filled=empty, decor.size=6.5thin}{v1,v2,v4}
				\fmfv{decor.shape=circle,decor.filled=full, decor.size=6.5thin}{v3}
			\end{fmfgraph*}
		}
		-\parbox{25mm}{
			\begin{fmfgraph*}(75,25)
				\fmfpen{0.5thin}
				\fmftop{o1,o2,o3}
				\fmfbottom{u1,u2,u3}
				\fmf{phantom,tension=100}{u1,dl,v1,o2}
				\fmf{plain}{dul,v1,o2}
				\fmf{phantom,tension=100}{u3,dr,v2,o2}
				\fmf{plain}{dur,v2,o2}
				\fmf{phantom}{u1,dul,u2,dur,u3}
				\fmf{plain}{dul,u2,dur}
				\fmf{phantom,tension=0.5}{dl,dum,dr}
				\fmf{phantom,tension=1}{dum,u2}
				\fmf{plain}{v2,dum}
				\fmf{plain}{v1,dum}
				\fmfv{decor.shape=circle,decor.filled=empty, decor.size=6.5thin}{v1,v2,u2}
			\end{fmfgraph*}
		}
		-\parbox{25mm}{
			\begin{fmfgraph*}(75,25)
				\fmfpen{0.5thin}
				\fmftop{o0,o1,o2,o3,o4,o5,o6}
				\fmfbottom{u0,u1,u2,u3,u4,u5,u6}
				\fmf{phantom}{u1,v1,o3}
				\fmf{plain}{v1,o3}
				\fmf{phantom}{o1,v1,u3}
				\fmf{plain}{v1,u3}
				\fmf{phantom}{u3,v2,o5}
				\fmf{plain}{u3,v2}
				\fmf{phantom}{o3,v2,u5}
				\fmf{plain}{o3,v2}
				\fmf{phantom}{u0,v1,dummy1,o4}
				\fmf{plain}{v1,dummy1}
				\fmf{phantom}{u4,dummy2,v1,o0}
				\fmf{plain}{dummy2,v1}
				\fmf{phantom}{u2,dummy2,v2,o6}
				\fmf{plain}{dummy2,v2}
				\fmf{phantom}{u6,v2,dummy1,o2}
				\fmf{plain}{v2,dummy1}
				\fmfv{decor.shape=circle,decor.filled=empty, decor.size=6.5thin}{v1,v2}
			\end{fmfgraph*}
		}\mkern-30mu,
	\end{eqnarray}
\end{fmffile}because the spectacles diagram cancels, as in \eqref{Glasses_diagrams_fourth_order_GamJ}.
This reproduces the result from \cite{Kuehn18_375004} obtained
from the first Legendre transform. There, the argument of the Legendre
transform is just $\boldsymbol{\m}$, necessitating inner derivatives
when differentiating the second-order contribution to the free energy
(the TAP term) to obtain $\frac{\partial\GamJ_{V}}{\partial\boldsymbol{m}}$,
which leads to equation (48) in \cite{Kuehn18_375004}. It contains
the spectacles diagram and the corresponding terms with $\frac{\partial\GamJ_{V}}{\partial\boldsymbol{m}}$-derivatives,
which taken together, with the cumulants of the Ising model inserted,
generates the same expression as the counter diagram for the ring.
This appeared quite mysterious in \cite{Kuehn18_375004} - however,
extending the Legendre transform to one involving the second-order
source term gives now a much more natural and shorter derivation.
A similar simplification of the Ising diagrammatics, actually characterizing
the diagrams to all orders, was already presented in \cite{Vasiliev74,Vasiliev75}\cite[sec. 6.3.4.]{vasiliev2019functional},
however, featuring a quite technical derivation without link to Legendre
transforms with respect to higher-oder source fields.

\subsubsection{The ring resummation in the Ising model}

The fact that for the Ising model Legendre transforms with respect
to arbitrarily many source fields are identical to the first Legendre
transform justifies the resummation of the ring diagrams expressed
by matrix products for the Ising model as performed in \cite{Sessak09,Cocco12_252}.
In particular, with this results, we can explain an observation of
Cocco and Monasson \cite[appendix B]{Cocco12_252}. They note that
a diagram depicting a term in their small-correlation expansion (the
middle one in the second column of their figure 26) contributes with
the same prefactor to the entropy as the corresponding term to what
they call the mean-field approximation of the entropy, our \eqref{Entropy_Gauss}.
The reason for this is not obvious in their framework (which is not
diagrammatic, but that one from \cite{Sessak09}; the diagrams are
just a depiction of terms derived by other means, as in Maillard et
al. \cite{Maillard19_113301}).

With our analysis, however, it is now apparent that this is neither
a coincidence nor special to the fourth order - rather, it is a consequence
of the fact that for the Ising model, \eqref{Entropy_Gauss} is the
exact resummation of all ring diagrams, including summations over
pairwise unequal indices (what is called loops in \cite{Cocco12_252}
and simple cycles in \cite{Maillard19_113301}) but also those with
 indices occuring multiple times.

\section{Discussion\label{sec:Discussion}}

In this manuscript, we have derived an extension of the diagrammatic
framework developed in \cite{Kuehn18_375004,Kuehn23_115001} to encompass
free energies as a function of the variances of a stochastic variable.
We have demonstrated that this leads to diagrammatic rules harmoniously
adding to those developed so far, which enable us to understand diverse
perturbative expansions derived by other means. In particular, it
allowed us to complete the proof of a result on a system characterized
by a rotationally invariant matrix, which so far has only been conjectured
\cite{Maillard19_113301}, and to derive, on a theoretically grounded
way, a practical approximation for the entropy of a complex system
of which only a limited amount of data is available \cite{Kuehn25a_arxiv}.
So while the present manuscript, as we believe, is a valid contribution
because it relates known results to each other, thereby enables a
 better understanding,  simplifies their derivation and completes
them by allowing statements about arbitrary orders, we have not derived
any result that has not been already intuited or could have been guessed
by ``common sense''. However, because our diagrammatic framework
indeed allows for comparatively simple and compact derivations of
approximations, we believe that it will allow to tackle several timely
problems that have been hard to treat because of technical difficulties
so far and that we briefly discuss in the following.

\subsection{Further applications}

The reason for which the free energy has been studied in \cite{Maillard19_113301}
is that it enables the derivation of message-passing algorithms \cite[chap. 14]{Mezard_Montanari09},
notably in their implementation as belief propagation \cite{Yedidia03_239},
an efficient way to compute the statistics of a system. Besides providing
a way to solve many problems in data science, it is also at the core
of their theoretical study \cite{Zdeborova16_453}. In a follow-up
work to \cite{Maillard19_113301} for example, Maillard et al. consider
the problem of matrix factorization: finding the matrices $F$ and
$X$ from the observation of a matrix $Y$ obeying some law 
\[
Y\sim P\left(Y|FX\right).
\]
In what is known as the Bayesian optimal setting, that is, in the
case that the probability distribution $P$ is known, as well as the
prior distributions of $F$ and $X$, this problem amounts to computing
the joint probability distribution of $Y$, $F$ and $X$, which requires
to compute the corresponding partition function - or, equivalently,
the free energy. For most cases, it is infeasible to achieve this
exactly. It is therefore a matter of ongoing research to find good
approximations for the free energy - and therefore to find the corresponding
message-passage algorithms - for the problem of matrix factorization
\cite[and references therein]{Kabashima16_4228,Maillard22_083301}.
This is particularly true for the extensive-rank case, in which $F$
and $X$ have a rank in the same order as their size, which is treated
in \cite{Maillard22_083301} by means of the Georges/Yedidia high-temperature
expansion \cite{georges1991expand}. However, similar as in \cite{Maillard19_113301},
an exhaustive exploration of this perturbative inroad to the problem
is prevented by the technical limits of the Georges/Yedidia expansion.
It is therefore a promising venue for future research to apply our
diagrammatic framework to this problem and related questions.

Applying our approach as a tool to compute entropies of undersampled
systems, its extension to higher-order interactions is a natural step
to do, which promises to have multiple applications \cite{Battiston20_1}.
Formally treating higher-order interactions in our framework is straight-forward,
however, we expect it to be challenging to derive useful approximations
for them going beyond the lowest order in perturbation theory, possibly
including resummations, in particular in case that Legendre transforms
are performed with respect to the additional interactions. Here, approximations
adapted to the concrete applications at hand could become necessary.

\subsection{Further theoretical development}

For deriving our diagrammatic framework, we have treated the perturbative
expansions in the interactions or covariances as formal power series,
that is, we have ignored the question of convergence. For the first
Legendre transform (with respect to only one source field), it has
been shown that this procedure can be justified by introducing the
Legendre transform as an algebraic operation on the ring of power
series \cite{Jackson08_arxiv,Jackson17_225201}. The idea there is
to use the fact that the derivatives of a function can be represented
by the derivatives of its Legendre transform; more precisely, that
this representation can be depicted by tree diagrams, in which the
vertices represent the derivatives of the Legendre transform \cite[sec. 12]{Helias20_970}.
This property is then used as the defining property of the Legendre
transform. Correspondingly, the coefficients of the Legendre-transformed
power series - and thereby the corresponding diagrammatic rules -
are derived using enumerative combinatorics \cite{Cameron17_book},
in particular by using it in the context of the Lagrange inversion
theorem, similar to (but more rigorous than) what we have suggested
in an earlier study \cite[sec. 2.4.2.]{Kuehn18_375004}. It would
be interesting to implement this program for the second Legendre transform
to adapt it for our purposes. The complication is that because in
the present case there is more than one source field, the foundational
equations for the diagrammatics are matrix-valued, in particular

\begin{fmffile}{InverseRelationHessians_diagrams}
	\fmfset{thin}{0.75pt}
	\fmfset{decor_size}{4mm}
	\begin{equation}		
		\fl\parbox{30mm}{
			\begin{fmfgraph*}(100,30)
				\fmfleft{l1}
				\fmfright{r1}
				\fmf{wiggly}{l1,v1,z1}
				\fmf{plain}{z1,c1,r1}
				\fmfv{d.s=circle, d.filled=shaded}{v1}
				\fmfv{d.s=circle, d.filled=empty}{c1}				
			\end{fmfgraph*}		
		} \mkern 27mu=  1.
		\label{eq:relation_scd_derivatives}
	\end{equation}		
\end{fmffile}We detail in \subsecref{Hessian_GamJ} how this translates into formulas.
For Gaussian theories, the corresponding matrices are diagonal (see
\eqref{Hessian_Gamma_W}) and we therefore expect it to be straight-forward
to generalize the approach of reverse engineering the Legendre transform
in terms of formal power series. However, in the general, non-Gaussian
case, \eqref{relation_scd_derivatives} is a true matrix equation,
which renders the generalization non-trivial. Nonetheless, this inroad
is promising because it would make our approach amenable to the formalism
of enumerative combinatorics, yielding more insights into the structure
of our diagrams. Possibly, this will reveal more cancellations and
therefore allow even more compact expressions. These tools could also
be leveraged in a more limited setting, focusing, for example, on
certain types of crossing identifications, which could be analyzed
by means of group theoretical methods to identify further potential
cancellations, as indicated in \subsecref{Cancellation_higher_orders}.

As a further technical observation, we note that our counter diagrams
seem closely related to what Vasiliev and Radzhabov call compensating
diagrams in their diagrammatics of the first Legendre transform for
the Ising model \cite{Vasiliev75}\cite[sec. 6.3.4.]{vasiliev2019functional}.
Their analysis, however, is not restricted to sub-diagrams attached
to the rest by one or two legs, as ours is (because counter diagrams
come from differentiating $\GamJ$ with respect to the one-point cumulants
$m_{i}$ or the two-point cumulants $v_{i}$). Also, their derivation
is very different from ours, in particular they do not refer to any
Legendre transform of order higher than $1$. Nonetheless, we reckon
that their result could be obtained in a more straight-forward way
by formally defining the complete Legendre form, that is, with respect
to all source fields of arbitrary order, as foreshadowed in \subsecref{Application_Ising}.
We leave these inroads of further developing diagrammatics for future
research. 

\ack{}{}

We thank Laura Foini for very helpful discussions, Claudia Merger
for very helpful discussions and a careful reading of this manuscript
and Jonas Stapmanns for advice during the review process.

We acknowledge funding by ANR-21-CE37-0024 NatNetNoise. This work
has been done within the framework of the PostGenAI@Paris project
and it has benefitted from financial support by the Agence Nationale
de la Recherche (ANR) with the reference ANR-23-IACL-0007. Our lab
(Institut de la Vision) is part of the DIM C-BRAINS, funded by the
Conseil Régional d\textquoteright Ile-de-France. 

\section{Appendix}

\subsection{General treatment of $\boldsymbol{m}$- and $\boldsymbol{v}$-dependence
of unperturbed cumulants\label{subsec:Hessian_GamJ}}

In this section, we will derive the Hessian of $\GamJ_{0}$. We will
omit the index $i$ in the following. We have, using $\frac{\partial W}{\partial j}=m$
for the underbraced expressions, 
\begin{eqnarray*}
\fl1=\frac{\partial m}{\partial m} & = & \frac{\partial j}{\partial m}\frac{\partial^{2}W}{\partial j^{2}}+\frac{\partial k}{\partial m}\frac{\partial^{2}W}{\partial k\partial j}=\frac{\partial}{\partial m}\left(\frac{\partial\Gamma}{\partial m}-2m\frac{\partial\Gamma}{\partial v}\right)\frac{\partial^{2}W}{\partial j^{2}}+\frac{\partial^{2}\Gamma}{\partial v\partial m}\frac{\partial^{2}W}{\partial k\partial j}\\
\fl & = & \left(\frac{\partial^{2}\Gamma}{\partial m^{2}}-2\frac{\partial\Gamma}{\partial v}-2m\frac{\partial^{2}\Gamma}{\partial m\partial v}\right)\frac{\partial^{2}W}{\partial j^{2}}+\frac{\partial^{2}\Gamma}{\partial v\partial m}\frac{\partial}{\partial j}\left(\frac{\partial^{2}W}{\partial j^{2}}+\left(\frac{\partial W}{\partial j}\right)^{2}\right)\\
\fl & = & \left(\frac{\partial^{2}\Gamma}{\partial m^{2}}-2\frac{\partial\Gamma}{\partial v}\right)\frac{\partial^{2}W}{\partial j^{2}}+\frac{\partial^{2}\Gamma}{\partial v\partial m}\frac{\partial^{3}W}{\partial j^{3}}\\
\fl &  & \underbrace{-2m\frac{\partial^{2}\Gamma}{\partial m\partial v}\frac{\partial^{2}W}{\partial j^{2}}+2\frac{\partial^{2}\Gamma}{\partial v\partial m}\frac{\partial W}{\partial j}\frac{\partial^{2}W}{\partial j^{2}}}_{=0}\\
\fl0=\frac{\partial m}{\partial v} & = & \frac{\partial j}{\partial v}\frac{\partial^{2}W}{\partial j^{2}}+\frac{\partial k}{\partial v}\frac{\partial^{2}W}{\partial k\partial j}=\frac{\partial}{\partial v}\left(\frac{\partial\Gamma}{\partial m}-2m\frac{\partial\Gamma}{\partial v}\right)\frac{\partial^{2}W}{\partial j^{2}}+\frac{\partial^{2}\Gamma}{\partial v^{2}}\frac{\partial^{2}W}{\partial k\partial j}\\
\fl & = & \left(\frac{\partial^{2}\Gamma}{\partial m\partial v}-2m\frac{\partial^{2}\Gamma}{\partial v^{2}}\right)\frac{\partial^{2}W}{\partial j^{2}}+\frac{\partial^{2}\Gamma}{\partial v^{2}}\frac{\partial}{\partial j}\left(\frac{\partial^{2}W}{\partial j^{2}}+\left(\frac{\partial W}{\partial j}\right)^{2}\right)\\
\fl & = & \frac{\partial^{2}\Gamma}{\partial m\partial v}\frac{\partial^{2}W}{\partial j^{2}}+\frac{\partial^{2}\Gamma}{\partial v^{2}}\frac{\partial^{3}W}{\partial j^{3}}\\
\fl &  & \underbrace{-2m\frac{\partial^{2}\Gamma}{\partial v^{2}}\frac{\partial^{2}W}{\partial j^{2}}+2\frac{\partial^{2}\Gamma}{\partial v^{2}}\frac{\partial^{2}W}{\partial j^{2}}\frac{\partial W}{\partial j}}_{=0}\\
\fl0=\frac{\partial v}{\partial m} & = & \frac{\partial j}{\partial m}\frac{\partial}{\partial j}\frac{\partial^{2}W}{\partial j^{2}}+\frac{\partial k}{\partial m}\frac{\partial}{\partial k}\frac{\partial^{2}W}{\partial j^{2}}=\frac{\partial}{\partial m}\left(\frac{\partial\Gamma}{\partial m}-2m\frac{\partial\Gamma}{\partial v}\right)\frac{\partial^{3}W}{\partial j^{3}}+\frac{\partial^{2}\Gamma}{\partial m\partial v}\frac{\partial^{2}}{\partial j^{2}}\left(\frac{\partial^{2}W}{\partial j^{2}}+\left(\frac{\partial W}{\partial j}\right)^{2}\right)\\
\fl & = & \left(\frac{\partial^{2}\Gamma}{\partial m^{2}}-2\frac{\partial\Gamma}{\partial v}\right)\frac{\partial^{3}W}{\partial j^{3}}+\frac{\partial^{2}\Gamma}{\partial m\partial v}\left(\frac{\partial^{4}W}{\partial j^{4}}+2\left(\frac{\partial^{2}W}{\partial j^{2}}\right)^{2}\right)\\
 &  & \underbrace{-2m\frac{\partial^{2}\Gamma}{\partial v\partial m}\frac{\partial^{3}W}{\partial j^{3}}+\frac{\partial^{2}\Gamma}{\partial m\partial v}2\frac{\partial W}{\partial j}\frac{\partial^{3}W}{\partial j^{3}}}_{=0}\\
\fl1=\frac{\partial v}{\partial v} & = & \frac{\partial j}{\partial v}\frac{\partial}{\partial j}\frac{\partial^{2}W}{\partial j^{2}}+\frac{\partial k}{\partial v}\frac{\partial}{\partial k}\frac{\partial^{2}W}{\partial j^{2}}=\frac{\partial}{\partial v}\left(\frac{\partial\Gamma}{\partial m}-2m\frac{\partial\Gamma}{\partial v}\right)\frac{\partial^{3}W}{\partial j^{3}}+\frac{\partial^{2}\Gamma}{\partial v^{2}}\frac{\partial^{2}}{\partial j^{2}}\left(\frac{\partial^{2}W}{\partial j^{2}}+\left(\frac{\partial W}{\partial j}\right)^{2}\right)\\
\fl & = & \frac{\partial^{2}\Gamma}{\partial m\partial v}\frac{\partial^{3}W}{\partial j^{3}}+\frac{\partial^{2}\Gamma}{\partial v^{2}}\left(\frac{\partial^{4}W}{\partial j^{4}}+2\left(\frac{\partial^{2}W}{\partial j^{2}}\right)^{2}\right)\\
 &  & \underbrace{-2m\frac{\partial^{2}\Gamma}{\partial v^{2}}\frac{\partial^{3}W}{\partial j^{3}}+\frac{\partial^{2}\Gamma}{\partial v^{2}}2\frac{\partial W}{\partial j}\frac{\partial^{2}W}{\partial j^{2}}}_{=0}.
\end{eqnarray*}
In matrix form this reads
\begin{equation}
\fl\left(\begin{array}{cc}
1 & 0\\
0 & 1
\end{array}\right)=\left(\begin{array}{cc}
\frac{\partial^{2}\Gamma}{\partial m^{2}}-2\frac{\partial\Gamma}{\partial v} & \frac{\partial^{2}\Gamma}{\partial m\partial v}\\
\frac{\partial^{2}\Gamma}{\partial m\partial v} & \frac{\partial^{2}\Gamma}{\partial v^{2}}
\end{array}\right)\left(\begin{array}{cc}
\frac{\partial^{2}W}{\partial j^{2}} & \frac{\partial^{3}W}{\partial j^{3}}\\
\frac{\partial^{3}W}{\partial j^{3}} & \left(\frac{\partial^{4}W}{\partial j^{4}}+2\left(\frac{\partial^{2}W}{\partial j^{2}}\right)^{2}\right)
\end{array}\right).\label{eq:Hessian_Gamma_W}
\end{equation}
After inverting this matrix, we obtain the inner derivatives as
\begin{eqnarray*}
\frac{\partial j}{\partial m} & = & \frac{\partial^{2}\Gamma}{\partial m^{2}}-2\frac{\partial\Gamma}{\partial v}-2m\frac{\partial^{2}\Gamma}{\partial m\partial v}\\
\frac{\partial k}{\partial m} & = & \frac{\partial^{2}\Gamma}{\partial m\partial v}\\
\frac{\partial j}{\partial v} & = & \frac{\partial^{2}\Gamma}{\partial m\partial v}-2m\frac{\partial^{2}\Gamma}{\partial v^{2}}\\
\frac{\partial k}{\partial v} & = & \frac{\partial^{2}\Gamma}{\partial v^{2}},
\end{eqnarray*}
that we would need to evaluate the additional diagrams coming about
by $\frac{\partial\GamJ}{\partial\boldsymbol{m}}$ and $\frac{\partial\GamJ}{\partial\boldsymbol{v}}$
vertices containing cumulants of order higher than $2$, which, however,
we have not discussed in detail in this manuscript because it was
not required for our analysis.

\end{document}